\documentclass[twocolumn]{aastex631}

\usepackage{amsmath}
\usepackage{makecell}
\usepackage[T1]{fontenc}

\shorttitle{Revising the Giant Planet Mass-Metallicity Relation}
\shortauthors{Chachan et al.}

\begin{document}

\title{Revising the Giant Planet Mass-Metallicity Relation:\\ Deciphering the Formation Sequence of Giant Planets}

\correspondingauthor{Yayaati Chachan}
\email{ychachan@ucsc.edu}

\author[0000-0003-1728-8269]{Yayaati Chachan}
\affiliation{Department of Astronomy and Astrophysics, University of California, Santa Cruz, CA 95064, USA}

\author[0000-0002-9843-4354]{Jonathan J. Fortney}
\affiliation{Department of Astronomy and Astrophysics, University of California, Santa Cruz, CA 95064, USA}

\author[0000-0003-3290-6758]{Kazumasa Ohno}
\affiliation{Division of Science, National Astronomical Observatory of Japan, 2-12-1 Osawa, Mitaka, Tokyo 181-8588, Japan}

\author[0000-0002-5113-8558]{Daniel Thorngren}
\affiliation{Department of Physics and Astronomy, Johns Hopkins University, Baltimore, MD 21210, USA}

\author[0000-0001-5061-0462]{Ruth Murray-Clay}
\affiliation{Department of Astronomy and Astrophysics, University of California, Santa Cruz, CA 95064, USA}

\begin{abstract}
The rate at which giant planets accumulate solids and gas is a critical component of planet formation models, yet it is extremely challenging to predict from first principles. Characterizing the heavy element (everything other than hydrogen and helium) content of giant planets provides important clues about their provenance. Using thermal evolution models with updated H-He EOS and atmospheric boundary condition that varies with envelope metallicity, we quantify the bulk heavy element content of 147 warm ($< 1000$ K) giant planets with well-measured masses and radii, more than tripling the sample size studied in \cite{Thorngren2016}. These measurements reveal that the population's heavy element mass follows the relation $M_{\rm Z} = M_{\rm core} + f_Z (M_{\rm p} - M_{\rm core})$, with $M_{\rm core} = 14.7^{+1.8}_{-1.6}$ Earth masses (M$_\oplus$), $f_Z = 0.09 \pm 0.01$, and an astrophysical scatter of $0.66 \pm 0.08 \times M_Z$. The classical core-accretion scenario ($Z_{\rm p} = 1$ at 10 M$_\oplus$ and $Z_{\rm p} = 0.5$ at 20 M$_\oplus$) is inconsistent with the population. At low planet masses ($<< 150$ M$_\oplus$), $M_{\rm Z} \sim M_{\rm core}$ and as a result, $Z_{\rm p} = M_{\rm Z} / M_{\rm p}$ declines linearly with $M_{\rm p}$. However, bulk metallicity does not continue to decline with planet mass and instead flattens out at $f_Z \sim 0.09$ ($\sim 7 \times$ solar metallicity). When normalized by stellar metallicity, $Z_{\rm p} / Z_\star$ flattens out at $3.3 \pm 0.5$ at high planet masses. This explicitly shows that giant planets continue to accrete material enriched in heavy elements during the gas accretion phase. 
\end{abstract}

\section{Introduction} \label{sec:intro}

Planets form in protoplanetary disks by accumulating solids and gas. Jupiter-like gas giant planets are some of the most challenging objects to explain, as they must accrete enormous amounts of mass prior to the dispersal of the gas disk, which happens on the scale of a few million years. Currently, the most widely accepted paradigm for planet formation is a bottom up process - the core accretion theory (\citealp{Mizuno1980, Stevenson1982, Pollack1996}, see \citealp{Ikoma2025} for a recent review) - in which planetary cores form first from solids in the form of mm-cm sized `pebbles' \citep{Ormel2010, Lambrechts2012} and/or $0.1- 100$ km sized planetesimals \citep{Kokubo1998, Rafikov2004, Goldreich2004a, Kobayashi2011}. Once a core becomes sufficiently massive, it begins to accrete a hydrogen-rich gaseous envelope. However, the gas’ accretion rate is regulated by its ability to cool, which in turn depends on the extent of simultaneous solid accretion as well as the `dustiness' (small grains that contribute significantly to opacity) of the accreted gas: the amount of solids accreted determines the timescale over which a planet accretes a massive gas envelope \citep{Ikoma2000, Rafikov2006, Piso2014, Lee2014}. Eventually, if the envelope becomes so massive (comparable to the core mass) that it undergoes thermodynamic collapse and cools rapidly, runaway gas accretion ensues and Jupiter-like giant planets are formed. The amount of solids accreted during this runaway stage is challenging to predict from first principles. Naively, one might assume that the solid-to-gas ratio of the accreted material matches the dust-to-gas ratio of the protoplanetary disk. However, dust in the disk grows to mm size \citep{Birnstiel2024}, turns into planetesimals \citep{Chiang2010}, and dynamically decouples from the gas. Numerous studies have investigated the accretion of planetesimals during the formation and runaway accretion of giant planets and found that the extent of accretion depends sensitively on the planetesimal-planet dynamical interactions, damping of dynamical excitation of planetesimals by nebular gas, size of the planetesimals, evolutionary tracks of the growing planets, and gap opening in the disk \citep{Guillot2000, Rafikov2004, Zhou2007, Shiraishi2008, Shibata2019, Eriksson2022}. Given the difficulty in predicting the relative amount of solids and gas accreted by giant planets and the influence of this relative accretion on the planets' formation and composition, it is extremely valuable to make empirical measurements of this quantity. 

An excellent observational probe of the relative amounts of solids and gas accreted by giant planets is their bulk heavy element content (also termed `metallicity') given that the accreted gas is primarily composed of hydrogen and helium and solids contains heavy elements. The bulk metallicity of a giant planet is encoded in its mass and radius. If a planet is smaller than an equivalent mass of H-He, then we can immediately surmise that it must contain heavy elements \citep{Zapolsky1969, Stevenson1982a}. This relation also depends on the thermal history of the planet as the radius of a planet declines slowly over Gyr timescales. With a well-characterized planet population with measured masses, radii, and ages, we can quantify the bulk metallicity of giant planets and study how bulk metallicity varies with planet mass. 

This work was pioneered by \cite{Guillot2006} for a small sample of inflated hot-Jupiters and for cooler uninflated warm-Jupiters by \cite{Miller2011}. Studying the cooler uninflated planets is advantageous as it avoids the confounding effect of the unknown inflation mechanism on the thermal evolution and bulk metallicity estimation of hot-Jupiters. \cite{Thorngren2016} later expanded the latter study to population-level study of giant planets with a sample size of 47 planets and found a power law relation between planet mass and bulk metallicity. They established that bulk metallicity declines with increasing planet mass but not as rapidly as $1 / M_{\rm p}$ and that the total metal mass ($Z_{\rm p} \times M_{\rm p}$) of giant planets increases with $M_{\rm p}$. However, it was difficult to discern any additional information about the dependence of bulk metallicity on planet mass due to the sample size and the precision of the available measurements. For example, we expect the power law to flatten out at low masses for which $Z_{\rm p} \rightarrow 1$. Similarly, at high planet masses, one might expect the bulk metallicity to flatten out rather than continue to decrease to arbitrary low values. The potential non-power law nature of the mass-metallicity relation was also hinted at in follow-up studies that used the \cite{Thorngren2016} sample and found that removing a fixed amount of metals ($\sim 20$ M$_\oplus$) from all planets decreased or removed the observed correlation between planet mass and metallicity \citep{Hasegawa2018, Muller2020}.

In this work, we expand on the seminal work of \cite{Thorngren2016} by revisiting the giant planet sample after nearly a decade. The planet sample is much larger and the precision of the measurements is significantly higher, thanks to the continuous efforts of the exoplanet community to characterize these planets in detail. We leverage these improvements in the planet sample to reveal additional features in the bulk metallicity distribution of the planet population and make new inferences about the planet mass-metallicity relation. As part of this process, we also perform numerous upgrades to our evolution models. This includes the use of a modern and up-to-date equation of state for H-He, incorporating the effect of temperature-dependence of density and entropy of metals, modeling the core as an adiabat that contributes to the planetary luminosity, and accounting for the effect of adding metals to the envelope on their cooling and thermal evolution via the atmospheric boundary condition. 

\begin{figure}
    \centering
    \includegraphics[width=\linewidth]{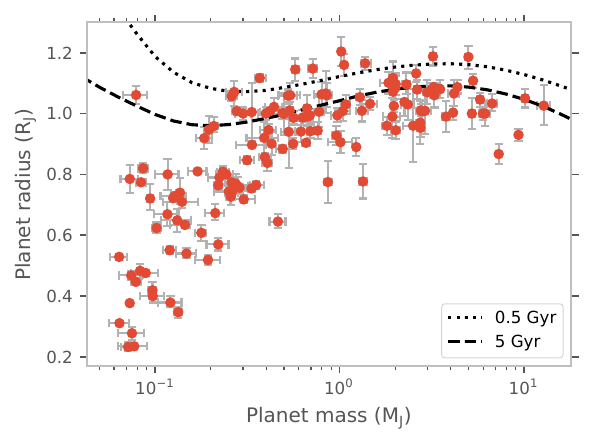}
    \caption{The radii and masses of the planets in our sample. The dotted and dashed lines show the mass-radius relation for pure H-He planets at 0.5 and 5 Gyr for an incident flux of $10^8$ erg s$^{-1}$ cm$^{-2}$.}
    \label{fig:planet_sample}
\end{figure}

\section{Planet sample}
We take the confirmed planet sample with both measured masses and radii from the Exoplanet Archive \citep{Christiansen2025} and make the following cuts to arrive at our sample. 
We limit ourselves to $20~M_\oplus - 20~M_{\rm Jup}$ and for equilibrium temperatures $T_{\rm eq} < 1000$ K. Planets hotter than this threshold are empirically known to be inflated but their inflation mechanism is not definitively established, which prevents us from using evolution models to estimate their bulk metal content \citep{Guillot2006, Thorngren2018}. We account for the effect of eccentricity on the time-averaged $T_{\rm eq}$ using Equation 15 of \cite{mendezEquilibriumTemperaturePlanets2017}. There are four circumbinary planets in our sample and for these, we calculate $T_{\rm eq}$ by using the summed flux of the two host stars with planet's semi-major axis measured from the binary's center of mass as the distance from the stars and assuming the planets have circular orbits. For planets with multiple data sets on the archive, we pick the data set that has the most precise mass and radius for a given planet. This often corresponds to the default data set on the archive but there for some planets, there are notable differences. Planets with radius and/or mass uncertainties that exceed 20\% of their measured value are excluded. The remaining planets typically have significantly lower uncertainties than this cutoff threshold. Figure~\ref{fig:planet_sample} shows the masses and radii of the planets in our sample as well as mass-radius curves for pure H-He objects at 0.5 and 5 Gyr with an incident flux of $10^8$ erg s$^{-1}$ cm$^{-2}$. Since the radius of giant planets is varies little with mass for a fixed composition, the mass-radius relation of planets in Figure~\ref{fig:planet_sample} immediately informs us that the bulk composition of giant planets varies with their mass.

\section{Methods} \label{sec:methods}

\subsection{Structure equations}
We solve the standard structure equations (hydrostatic equilibrium, mass conservation, conservation of energy) for giant planets using the relaxation method
\begin{equation}
    \frac{{\rm d} P}{{\rm d} m} = - \frac{G m}{4 \pi r^4}
\end{equation}
\begin{equation}
    \frac{{\rm d} r}{{\rm d} m} = \frac{1}{4 \pi r^2 \rho}
\end{equation}
\begin{equation}
    \frac{{\rm d} L}{{\rm d} m} = - T \frac{{\rm d} s_{\rm env}}{{\rm d} t} - T \frac{{\rm d} s_{\rm core}}{{\rm d} t}
\end{equation}
Here, $P$, $\rho$, $T$, $s_{\rm env}$ and $s_{\rm core}$ are the pressure, density, temperature, and specific entropy of the envelope and the core, respectively. We solve the equations over a mass grid with $m$ representing the mass coordinate and $r$ representing the corresponding radius coordinate. The planet loses heat due to its intrinsic luminosity $L$, which reduces its entropy over time.

Given a simple initial structure model guess, we iterate the solution until the difference between successive iterations becomes smaller than 0.1\%. The entire interior is assumed to follow an adiabat that cools over time. Naively, one would expect any non-adiabaticity to make the interior hotter and increase the amount of metals needed to match a planet's radius for a given mass and age. However, this is only true if the entropy of the outer convective envelopes are assumed to be the same, which is usually not the case for models with different initial metal distributions at a given age. In general, there are many arbitrary choices that need to be made about the initial metal profile and the initial entropy profile of planetary interiors that make it difficult to extract a generally applicable estimate of the difference between adiabatic and non-adiabatic models. Recent work shows that at the older ages ($\gtrsim 1$ Gyr) that are of relevance to our work, the radius of a planet with a compositional gradient is very similar to a planet with an adiabatic interior \citep[][]{Knierim2025, TejadaArevalo2025}. Figure E.1 in \cite{Knierim2025} shows that a homogeneously mixed planet has a radius that is nearly identical to one with compositional gradients, especially when the latter starts out with a higher initial entropy (see also Figure~3 in \citealt{TejadaArevalo2025}). For lower entropies, the difference in the radii at $\gtrsim 1$ Gyr is small — on the order $\lesssim 2$\%, which is within the measurement uncertainties of planetary radii and comparable to the uncertainty introduced by modeling choices even for fully adiabatic models. Accounting for compositional gradients should therefore have a small effect on the inferred bulk metallicities.

The core's structure is modeled with the pure-metal adiabat corresponding to the temperature and pressure at the core-envelope boundary. Details of the equation of state used for relating temperature to pressure and entropy ($T \equiv T(P, S)$) and the density to pressure and temperature ($\rho \equiv \rho(P, T)$) are discussed below (\S~\ref{sec:eos}). The intrinsic luminosity of the planet is determined by the atmospheric boundary that controls the rate at which the interior cools (\S~\ref{sec:atm_boundary}). We include the cooling of both the envelope and the core in our luminosity evolution and assume that the temperature at the core-envelope boundary is continuous. That is, decreasing envelope entropy causes a corresponding decline in the core's entropy and these are related by:
\begin{equation}
    \frac{{\rm d} s_{\rm core}}{{\rm d} t} = \frac{{\rm d} s_{\rm core}}{{\rm d} s_{\rm env}}  \frac{{\rm d} s_{\rm env}}{{\rm d} t},   
\end{equation}
where
\begin{equation}
    \frac{{\rm d} s_{\rm core}}{{\rm d} s_{\rm env}} = \frac{\partial s_{\rm core}}{\partial {\rm log}P}  \frac{\partial {\rm log}P}{\partial s_{\rm env}} + \frac{\partial s_{\rm core}}{\partial {\rm log}T}  \frac{\partial {\rm log}T}{\partial s_{\rm env}}.
\end{equation}
The derivatives with respect to log$P$ and log$T$ are readily calculable from the equation of state.

For the planet's transit radius, we also add a contribution from an isothermal atmospheric layer (at the planet's $T_{\rm eq}$) between a pressure $P_{\rm tr}$ and 10 mbars. $P_{\rm tr}$ is either 10 bars (the upper boundary of the interior models) or the pressure at which the envelope's temperature $= T_{\rm eq}$, whichever is higher. We also account for the effect of metallicity on the mean molecular weight of the isothermal layer assuming $\mu_Z = 18$. Although this is a simplification of the atmosphere's temperature structure and contribution to the planet's radius, we find the difference between the radius predicted from this method and atmosphere models is negligible, except at very low gravities that correspond to extremely young planetary ages. 

\subsection{Equation of state}
\label{sec:eos}

For the hydrogen-helium mixture, we utilize the equation of state (EOS) from \cite{chabrierNewEquationState2019, chabrierNewEquationState2021} for a He mass fraction $Y = 0.275$. We add the entropy contribution of the nuclear spin $s_{\rm spin} = {\rm ln (2 s + 1)}$ k$_{\rm B}$/baryon for H ($s = 1/2$, $s = 0$ for He) to bring the entropy in alignment with \cite{Saumon1995}. The \cite{chabrierNewEquationState2021} EOS includes corrections that arise from the non-ideal nature of the H-He mixture. These corrections are calculated for $Y = 0.245$ \citep{Militzer2013}, which is lower than our adopted $Y$. However, the error due to this difference is negligibly small \citep{Howard2023b}. 

We use a water EOS from \cite{Mazevet2019} to account for all metals (see \S~\ref{sec:evol_model_comparison} for the effect of choice on the bulk metallicity). The additive volume law is used to calculate EOS for mixtures of H-He and water. For each metallicity, we create a new EOS table on the fly and calculate the density and entropy as a function of pressure and temperature. The entropy, pressure, and temperature relation is inverted to obtain temperature as a function of entropy and pressure. The specific entropy of a mixture declines with increasing metallicity as adding metals significantly reduces the number of constituent atoms/molecules (metals add significantly more mass per atom than H-He). Since the H-He and water EOS do not provide information about the extent of dissociation and ionization, we cannot calculate the mixing entropy that arises from mixing water and H-He (the entropy of mixing arising from dissociation of molecular H or its ionization is accounted for in the H-He EOS and similarly for water). We make a simple estimate of it in Appendix \ref{sec:S_mix_estimate} to quantify the extent to which it would affect our adiabats.

\subsection{Atmospheric boundary condition}
\label{sec:atm_boundary}

Planetary cooling is governed by the outer boundary of the planet, i.e., its atmosphere. The atmosphere controls how much heat escapes from the interior of the planet \citep{Fortney2007}. In \cite{Thorngren2016}, a solar metallicity atmosphere enshrouded all the planets regardless of their envelope metallicity. In this work, we update the atmospheric metallicity in tandem with the envelope metallicity to make the models more consistent. Planetary metallicity has a strong effect on the outgoing heat flux of the planet as it controls the atmospheric opacity at the radiative-convective boundary. To do this, we use a self-consistent non-gray radiative-convective equilibrium atmospheric model \citep{Fortney2005, Fortney2020, Ohno2023} to create a new grid of atmospheric models for a range of atmospheric metallicities (log ($Z$/solar) $\in [0, 2]$ with increments of 0.5), surface gravities (log ($g$/cm s$^{-2}$) $\in [1,5]$ with increments of 1), incident fluxes (log ($F_{\rm inc}$/erg s$^{-1}$ cm$^{-2}$) $\in [6,9]$ with increments of 1), and internal heat fluxes ($T_{\rm int} = 24 \times 10^{i/5}$ K, for $i \in [0, 11]$ with $i$ incremented by 1). This extends the boundary conditions used in \cite{Fortney2007} to metallicities higher than $1 \times$ solar, while also updating all atomic and molecular opacities.

We find the interior entropy that corresponds to each atmospheric profile by matching the interior temperatures from the EOS (for the corresponding atmosphere's metallicity) and the atmosphere at a pressure of 1 kbar. When the atmospheric profile is truncated at a lower pressure, we match the temperature at 10 bars. Slight differences in the adiabatic gradient of the our atmosphere and interior models can lead to a difference in the entropy calculated from different pressures. We shift the entropies inferred at 10 bar to match the values inferred at 1 kbar if the difference between the inferred values is $\geq 0.1 \; k_{\rm B}/$atom. For very low $T_{\rm int}$ and high log $g$ models (an uncommon occurrence), the radiative-convective boundary can be even deeper than 1 kbar. We use the relatively steady value of $\partial {\rm log}(T_{\rm int}) / \partial s$ to linearly extrapolate ${\rm log}(T_{\rm int})$ and correct the entropy values in this region of the parameter space. Linking the intrinsic heat flux to the envelope entropy allows us calculate $T_{\rm int}$ as a function of atmospheric metallicity, surface gravity, incident flux, and envelope entropy. For this mapping and interpolation during calculations, we find that it is advantageous to use entropy in terms of $k_{\rm B}/$atom as it keeps the entropy's grid dynamic range consistent with increasing atmospheric metallicity. A planet's envelope metallicity may exceed $100 \times$ solar and during brief phases of planetary evolution, a planet's properties may lie outside this grid. Beyond the range of our grid, we clip the input values to the edge of the atmosphere grid.

\begin{figure}
    \centering
    \includegraphics[width=\linewidth]{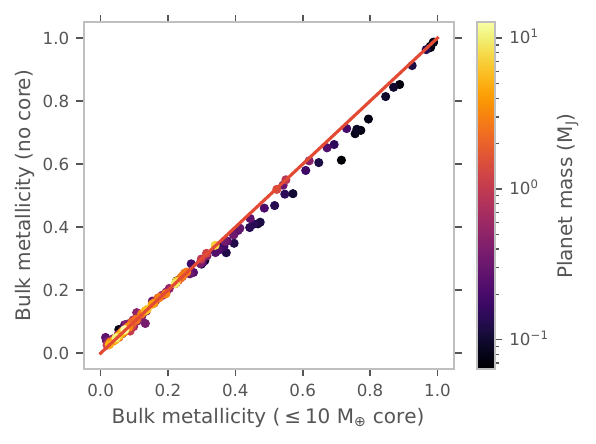}
    \caption{Median estimates of the bulk metallicity for our planet sample assuming the metals are distributed uniformly throughout the planet (y-axis) and our fiducial choice (x-axis, $M_{\rm Z} \leq 10 \, M_\oplus$ in the core, rest of the metals in the envelope).}
    \label{fig:no-core-compare}
\end{figure}

\subsection{Bulk metallicity estimation}

For each planet in our sample, we use the modeled planetary radius that depends on the planet's bulk metallicity to fit the observed radius given its mass, insolation, and age. 
Normal priors are placed on the planet's mass and on its age when age constraints are available. For planets without any uncertainties on their ages, we adopt an agnostic uniform prior $\mathcal{U}(0.5,10)$ Gyr for their age. The fits are performed using nested sampling with the open source code \texttt{dynesty} \citep{Speagle2020}.

The initial entropy is chosen to be either 18 $k_{\rm B}/$atom or the highest allowable by the combined metal and hydrogen-helium equations of state, whichever is smaller. Since all reasonable initial entropies converge onto the same evolution track within a few 100 Myrs \citep{marleyLuminosityYoungJupiters2007}, our results are not sensitive to this choice for older planets. We remove planets younger than 0.5 Gyr from our sample since their radius evolution depends on the choice of initial entropy. This leaves a total of 147 planets in our sample.

\begin{figure*}
    \centering
    \includegraphics[width=\linewidth]{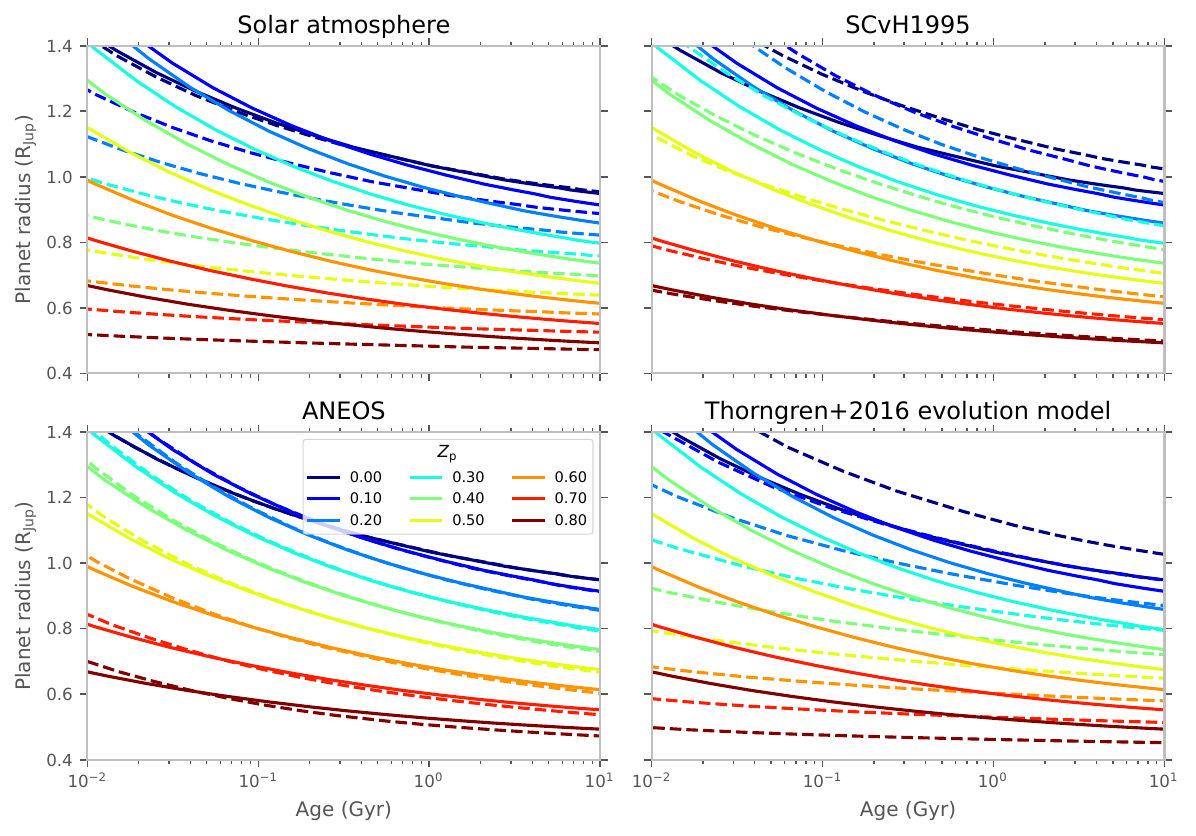}
    \caption{Evolution models for a 0.3 M$_{\rm Jup}$ planet for different bulk metallicities assuming the metals are uniformly distributed inside the planet (i.e., no central core). The solid lines correspond to our updated evolution model. The models corresponding to the dashed lines are indicated in the panel titles. In the top left panel, we compare with models that use a solar metallicity atmosphere but the same EOS for H-He and metals as our model. In the top right panel, we compare with models that use our adopted metal EOS and atmosphere models but with the SCvH EOS for H-He. In the bottom left panel, we compare with models that use ANEOS for metals. In the bottom right panel, we compare our evolution models with those calculated as in \cite{Thorngren2016}.}
    \label{fig:saturn_evol_comparison}
\end{figure*}

\begin{figure*}
    \centering
    \includegraphics[width=\linewidth]{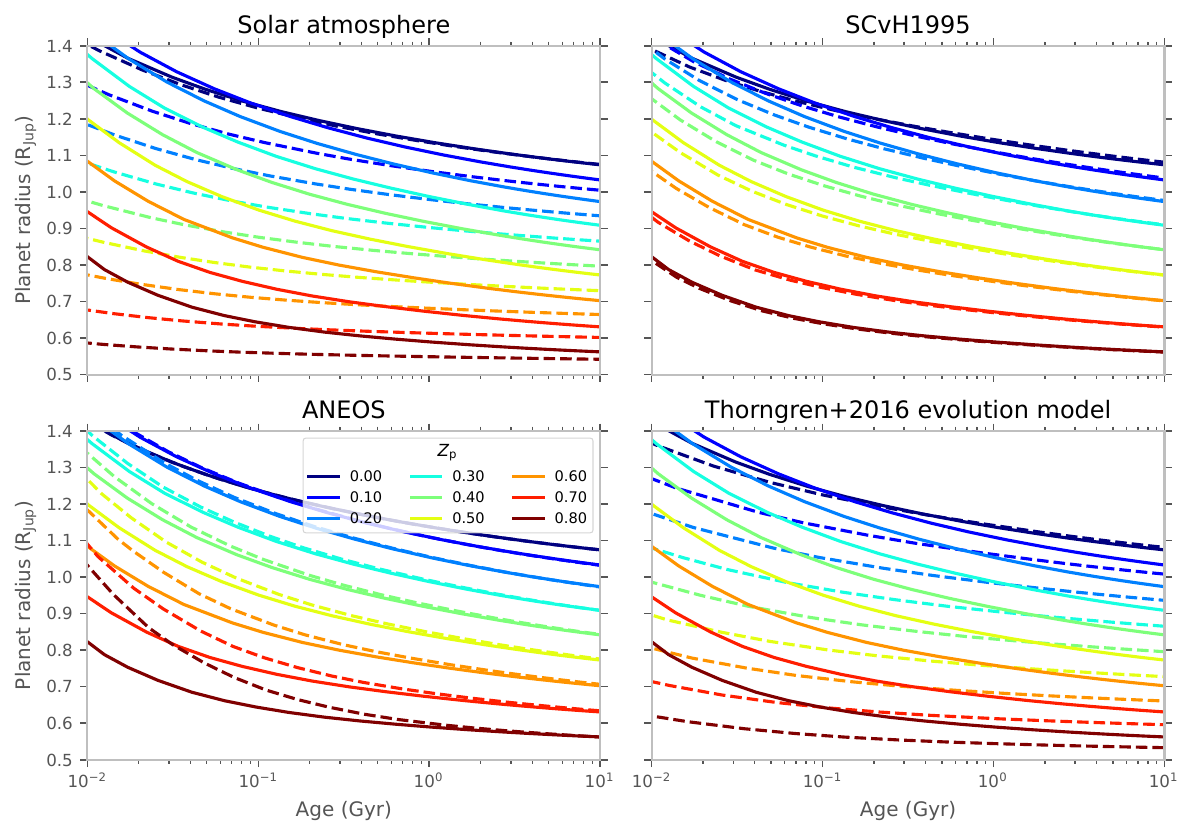}
    \caption{Evolution models for a 3 M$_{\rm Jup}$ planet for different bulk metallicities assuming the metals are uniformly distributed inside the planet. The solid lines correspond to our updated evolution model. The models corresponding to the dashed lines are indicated in the panel titles. See Figure~\ref{fig:saturn_evol_comparison} caption for more details.}
    \label{fig:superjup_evol_comparison}
\end{figure*}

To zeroth order, a giant planet's radius is insensitive to how the metals are distributed in the planet. However, small differences in the modeled radius creep in due to the compressibility of H-He. In this work, we incorporate all metals into a core until metal mass reaches $10~M_\oplus$ and put any additional metals uniformly in the envelope. This is a reasonable approximation for metal distribution inside planets in the core accretion paradigm. We quantify the effect of changing metal distribution in the interior on the bulk metallicity estimates in Figure~\ref{fig:no-core-compare}. Putting in all the metals in the core or distributing all the metals uniformly has a small effect on the estimated metallicity. Distributing the metals throughout the planet typically produces a smaller planet at the same metallicity due to the compressibility of H-He, even though added metals in the envelope would slow planetary cooling and tend to keep the planet larger. Bulk metallicity estimates differ by more than 1$\sigma$ only for 9 low mass planets ($\leq 0.125$ M$_{\rm Jup}$), all of which have high metallicities and metal masses $> 10$ M$_\oplus$. The absolute difference in the predicted metal mass for these planets is merely $0.8 - 2.2$ M$_\oplus$. In reality, we do not expect planets in this mass range to be fully mixed and devoid of a core. The choice of metal distribution leads to differences that are much smaller than the astrophysical scatter in our population level models (\S~\ref{sec:Mp_Zp_relation}) and can therefore be neglected.

\subsection{Comparison with previous model}
\label{sec:evol_model_comparison}

Given the numerous updates to our evolution models, the H-He EOS, metal EOS (including the effects of temperature and entropy), and the atmospheric boundary conditions, we compare the effect of these changes for a 0.3 M$_{\rm Jup}$ planet (Figure~\ref{fig:saturn_evol_comparison}) and 3 M$_{\rm Jup}$ planet (Figure~\ref{fig:superjup_evol_comparison}). The results we note below agree with previous studies that investigated the effects of various assumptions in planetary interior modeling on their evolution and bulk metallicity estimation \citep[e.g., ][]{Muller2023, Howard2025}. In the top left panel, we compare to models that use a solar metallicity atmosphere with our default EOS for H-He and metals. In the top right panel, we swap out the CMS H-He EOS with the SCvH EOS while keeping the metal EOS and the atmosphere grid the same. The entropy mapping in the atmosphere grid is recalculated using SCvH instead of CMS for this comparison. In the bottom left panel, we use ANEOS for metals with CMS EOS for H-He and our updated atmospheric boundary condition. In the bottom right panel, we compare our evolution models with those of \cite{Thorngren2016} that differ from our models in all these three respects. 

In the top left panels, planets with the updated atmosphere grids tend to be larger at all ages. This is a result of delayed cooling due to added opacity from the enhanced metallicity of the atmosphere, which is not accounted for by the solar metallicity atmosphere grid. The evolution curves for the pure H-He planet ($Z_{\rm p} = 0$) lie essentially on top of each other especially $\gtrsim 1$ Gyr, with some deviations due to small differences in the atmosphere grids at the lowest metallicity value ($1 \times$ solar). The difference between the models is largest at intermediate metallicities. This is because for low values of $Z_{\rm p}$, the effect of enhanced atmospheric metallicity is not as strong on the evolution of a planet. At very high values of $Z_{\rm p}$, the difference between the two models declines because the effect of increased density contribution from putting so much metals in the planet dominates over the delayed cooling from an enhanced atmospheric metallicity. 

\begin{figure*}
    \centering
    \includegraphics[width=\linewidth]{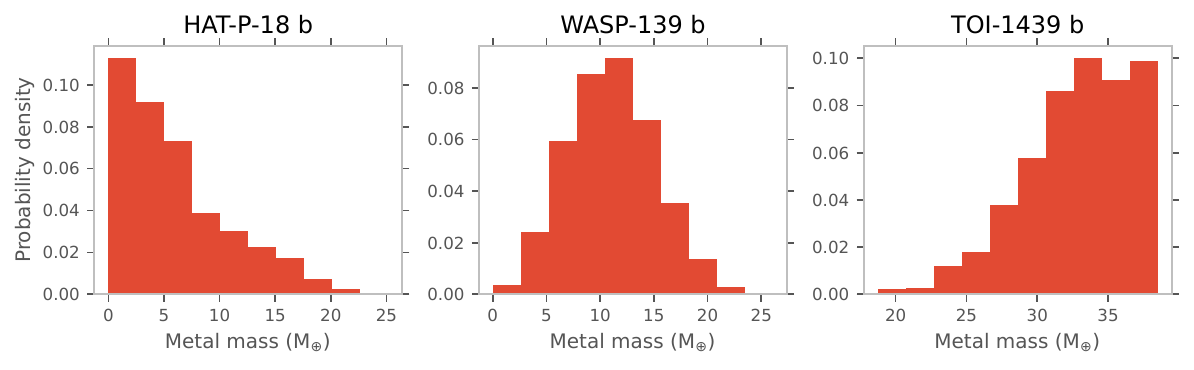}
    \caption{The probability distribution of the calculated metal mass for a sample of three planets. For HAT-P-18 b, the distribution provides an upper limit on its metal mass. TOI-1439 b's distribution is truncated at its total planet mass while WASP-139 b's metal distribution is nearly Gaussian. For planets such as HAT-P-18 b and TOI-1439 b, we fit a truncated normal distribution to the metal distribution and use it to calculate the model likelihood.}
    \label{fig:Mz_distribution_example}
\end{figure*}

The bottom left panels show the effect of swapping out the \cite{Mazevet2019} EOS for water with the $T=0$ EOS for a 50\%-50\% mixture of rock and ice. ANEOS is typically denser than the \cite{Mazevet2019} EOS. However the effect on planetary evolution is muted due to two reasons. Firstly, planets built with ANEOS only use H-He entropy for adiabats; these adiabats are steeper than the adiabats obtained by mixing H-He with \cite{Mazevet2019} EOS. As a result, the interior of a planet built with \cite{Mazevet2019} EOS is cooler, which increases the interior's density and partly compensates for the higher density of ANEOS. This difference would not be as important if a temperature and entropy dependent EOS is used for rocks, which should have cooler adiabats and produce slightly denser planets. Secondly, the larger size of planets built with \cite{Mazevet2019} EOS at early ages leads to rapid cooling and contraction, which can even lead them to be smaller than the planets built with ANEOS at $\lesssim 1$ Gyr (e.g., for a 3 M$_{\rm Jup}$ planet in Figure~\ref{fig:superjup_evol_comparison}). We run additional models with our default setup (10 M$_\oplus$ core and additional metals in the envelope) for 0.1, 1, and 10 M$_{\rm Jup}$ planets and bulk metallicities in the range of $0.1 - 0.9$. By comparing isochrones, we find that the bulk metallicities that produce the same planet radius for these two metal EOS choices differ by merely $\sim 0.01$ for $Z_{\rm p} = 0.1$ and $\sim 0.04$ for $Z_{\rm p} = 0.9$ for a 0.1 M$_{\rm Jup}$ planet.

In the top right panels, we show the effects of changing the H-He EOS on planet evolution curves, in particular due to the lower density of SCvH EOS in the $0.1 - 10$ Mbar region \citep{chabrierNewEquationState2019, chabrierNewEquationState2021, Howard2023a}. This effect is especially important for low mass planets for which much of their planet mass is present in this pressure range. For 0.3 M$_{\rm Jup}$, planets built with CMS EOS are significantly smaller than those built with SCvH EOS. The difference declines as the bulk metallicity is increased, as expected from increasingly important contributions to the density from the metals. However, for a 3 M$_{\rm Jup}$ planet, the evolution curves do not differ significantly because most of the mass is present at pressures $> 10$ Mbar, where SCvH and CMS do not differ. At earlier ages, the SCvH planets are even marginally smaller because their larger size very early on leads to rapid cooling (higher luminosity) and decrease in entropy.

The bottom right panels show the cumulative effect of all the differences in the evolution models of \cite{Thorngren2016} and this work. In general, the planets in the updated evolution models are larger with the $Z_p$ dependent behavior similar to the one described above the right most panel. At lowest metallicities, the updated evolution tracks produce smaller planets, especially in the Saturn mass regime due to the higher density of H-He in the CMS EOS.

\section{Planet Bulk metallicities}
\label{sec:bulk_metallicity_estimate}

\begin{figure*}
    \centering
    \includegraphics[width=0.49\linewidth]{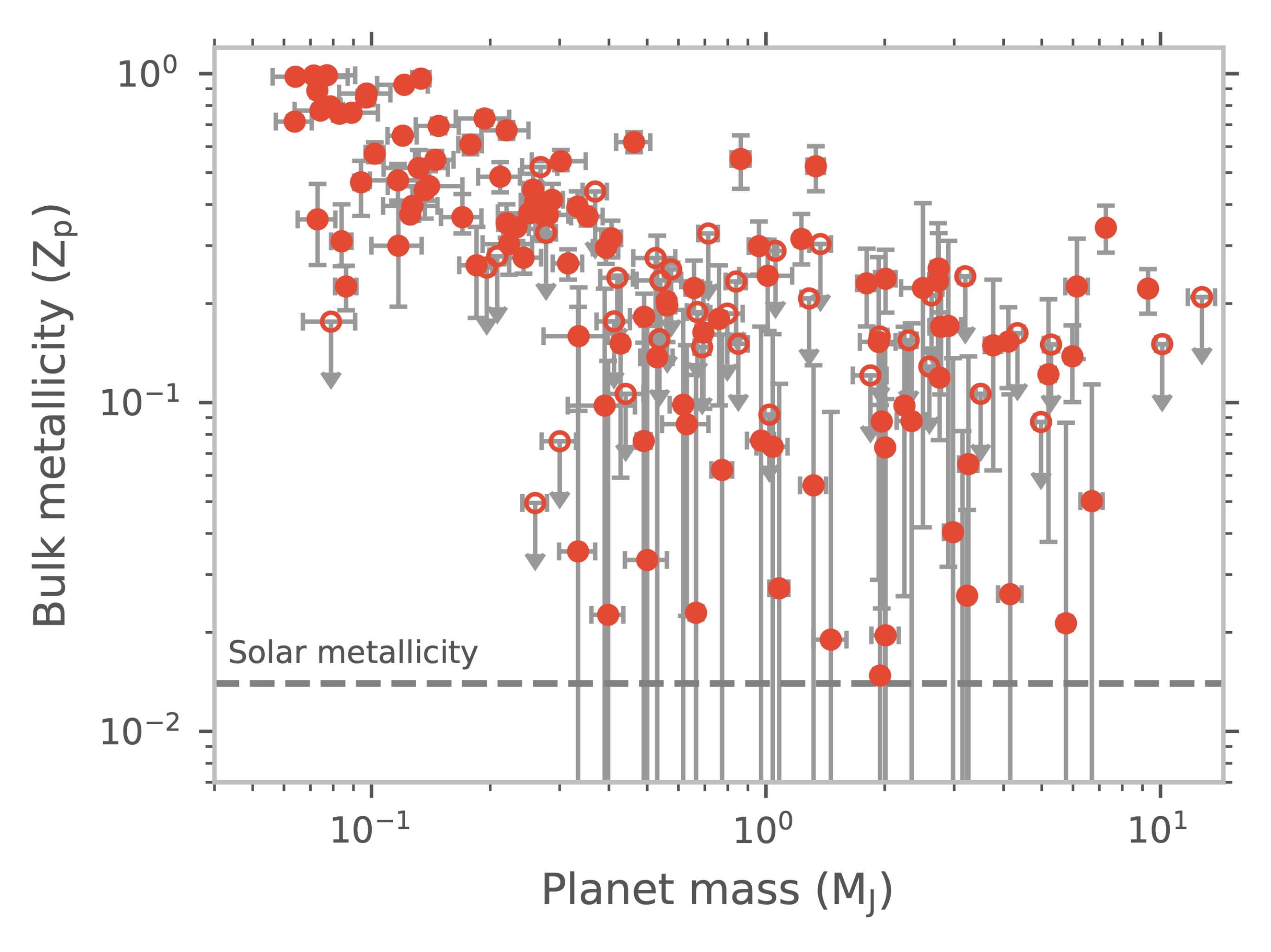}
    \includegraphics[width=0.49\linewidth]{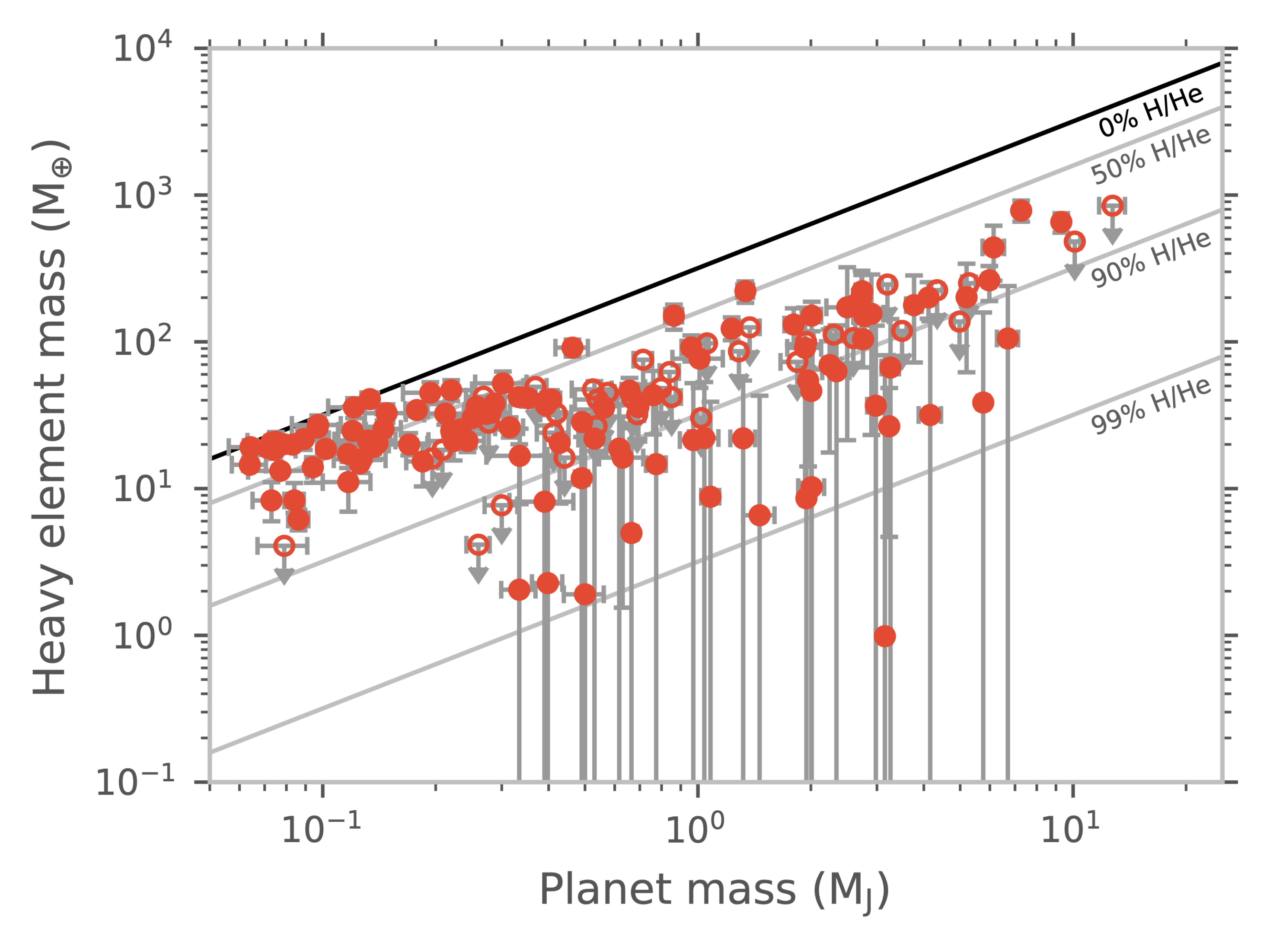}
    \caption{The bulk metallicity and metal mass plotted against planet mass for our sample. The dashed line in the left panel marks solar metallicity ($Z_\odot = 0.014$). Solids lines in the right panel show metal mass as a function of planet mass for a range of metallicities.}
    \label{fig:planet_metallicity}
\end{figure*}

Figure~\ref{fig:Mz_distribution_example} shows the metal mass posteriors for three planets in our sample and Figure~\ref{fig:planet_metallicity} shows the bulk metallicity $Z_{\rm p}$ and metal mass $M_{Z}$ (product of $Z_{\rm p}$ and $M_{\rm p}$) for our entire planet sample. For planets with $Z_{\rm p}$ within $3\sigma$ of 0 or 1, we fit a truncated normal distribution to the posterior to estimate its peak and standard deviation (with a similar procedure for $M_{Z}$ when it is within $3\sigma$ of 0 or $M_{\rm p}$, e.g., HAT-P-18 b and TOI-1439 b in Figure~\ref{fig:Mz_distribution_example}). There are 85 such planets in our sample. Of these, the $Z_{\rm p}$ posterior of 38 planets peaks at $Z_{\rm p} = 0$. For these planets, we plot the 95\% confidence interval upper limit in Figure~\ref{fig:planet_metallicity}.

\begin{figure*}
    \centering
    \includegraphics[width=0.75\linewidth]{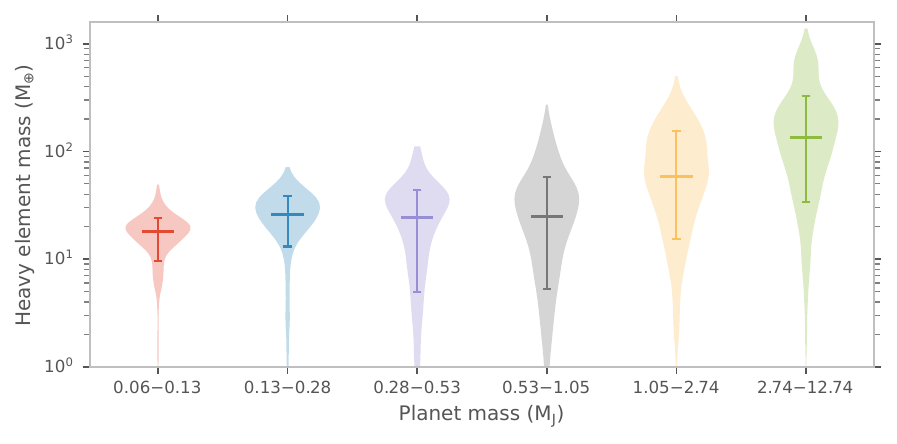}
    \includegraphics[width=0.75\linewidth]{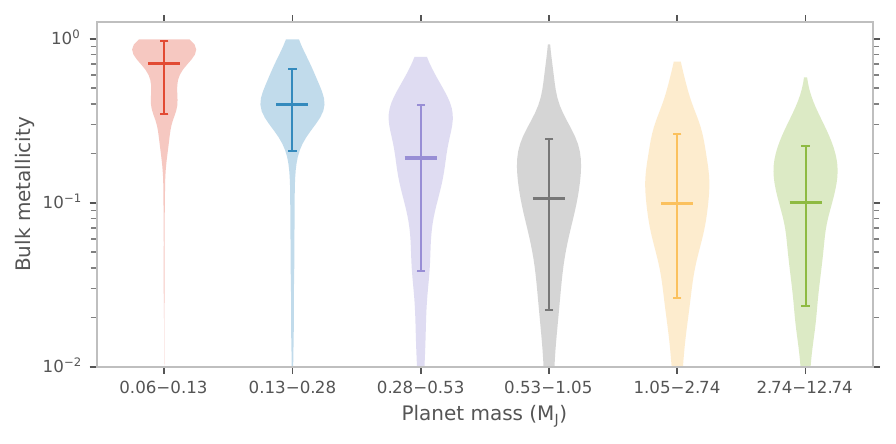}
    \caption{Violin plots for the metal mass and bulk metallicity of planets in our sample. The sample is split into six nearly equally sized bins. We take 100 samples for the bulk metallicity and metal mass for each planet and pool these samples together for all the planets in the mass bin to account for the individual- and population-level uncertainties in these quantities. The horizontal bar indicates the median of the population in each mass bin and the vertical line with caps indicates the 16th and 84th percentile of the population.}
    \label{fig:violinplots}
\end{figure*}

Our expanded sample of planets reveals many interesting features in the metal content of close-in giant planets at a population level. To elucidate these features, we show a violin plot for the metal mass and bulk metallicity in Figure~\ref{fig:violinplots}. The planet population is split into six bins with nearly equal number of planets in each bin. We draw 100 samples for $M_{Z}$ or $Z_{\rm p}$ for each planet and combine them together for all the planets in a given mass bin to obtain the distribution used to make the violin plots. This allows us to include both the individual- and population-uncertainty in our distribution. Firstly, essentially all of the sub-Saturns ($M_{\rm p} < 0.3~{\rm M_{J}}$) in our sample are clustered together at high metallicities and have $Z_{\rm p} \gtrsim 0.2$. This is not due to an observational bias as lower metallicity planets would be {\it larger} in size and {\it easier} to detect. Secondly, the bulk metallicity first drops below 0.1 in the $0.3-0.5$ M$_{\rm Jup}$ mass range. In the $0.3 - 1$ M$_{\rm Jup}$ range, a larger fraction of planets have metal mass $\lesssim 10$ M$_\oplus$ compared to planets less massive than $0.3$ M$_{\rm Jup}$. Thirdly, there are no clear signs of a strong decline in $Z_{\rm p}$ with increasing planet mass beyond $\sim 0.5~{\rm M_{J}}$. The metallicity distribution in this mass range instead seems to be roughly uniform with some (likely astrophysical) scatter. This leads to a significant rise in the metal mass with increasing planet mass. Fourthly, the median bulk metallicity in this mass range is elevated compared to the metallicity value for the Sun ($Z_\odot = 0.014$). We discuss other interesting sub-populations below.

\textit{Metal rich high mass planets:} A non-negligible fraction of massive giant planets ($\sim 0.5 - 10~{\rm M_{J}}$) have high metallicities that exceed 0.1 and even approach values as high as 0.5. The three most notable planets in the mass range $0.3 - 1$ M$_{\rm Jup}$ with well-constrained metallicities that are elevated relative to the population: TIC 13927066 b ($0.463$ M$_{\rm Jup}$, $Z_{\rm p} = 0.62 \pm 0.04$), WASP-59 b ($0.863$ M$_{\rm Jup}$, $Z_{\rm p} = 0.55 \pm 0.10$), and HATS-17 b ($1.34$ M$_{\rm Jup}$, $Z_{\rm p} = 0.52 \pm 0.08$). Beyond 5 M$_{\rm Jup}$, three planets have median metallicities $>0.2$ are: TIC 279401253 b ($6.14$ M$_{\rm Jup}$, $Z_{\rm p} = 0.23 \pm 0.08$), HAT-P-20 b ($7.27$ M$_{\rm Jup}$, $Z_{\rm p} = 0.34 \pm 0.06$), TOI-2373 b ($9.30$ M$_{\rm Jup}$, $Z_{\rm p} = 0.22 \pm 0.04$). Such metallicities imply that enormous amounts of metals ($\gtrsim 100~M_\oplus$) are sequestered in these planets' interiors. HAT-P-20 b with $M_Z = 784^{+133}_{-125}$ M$_\oplus$ has the highest well constrained metal mass in the planet sample.

\begin{figure*}
    \centering
    \includegraphics[width=0.55\linewidth]{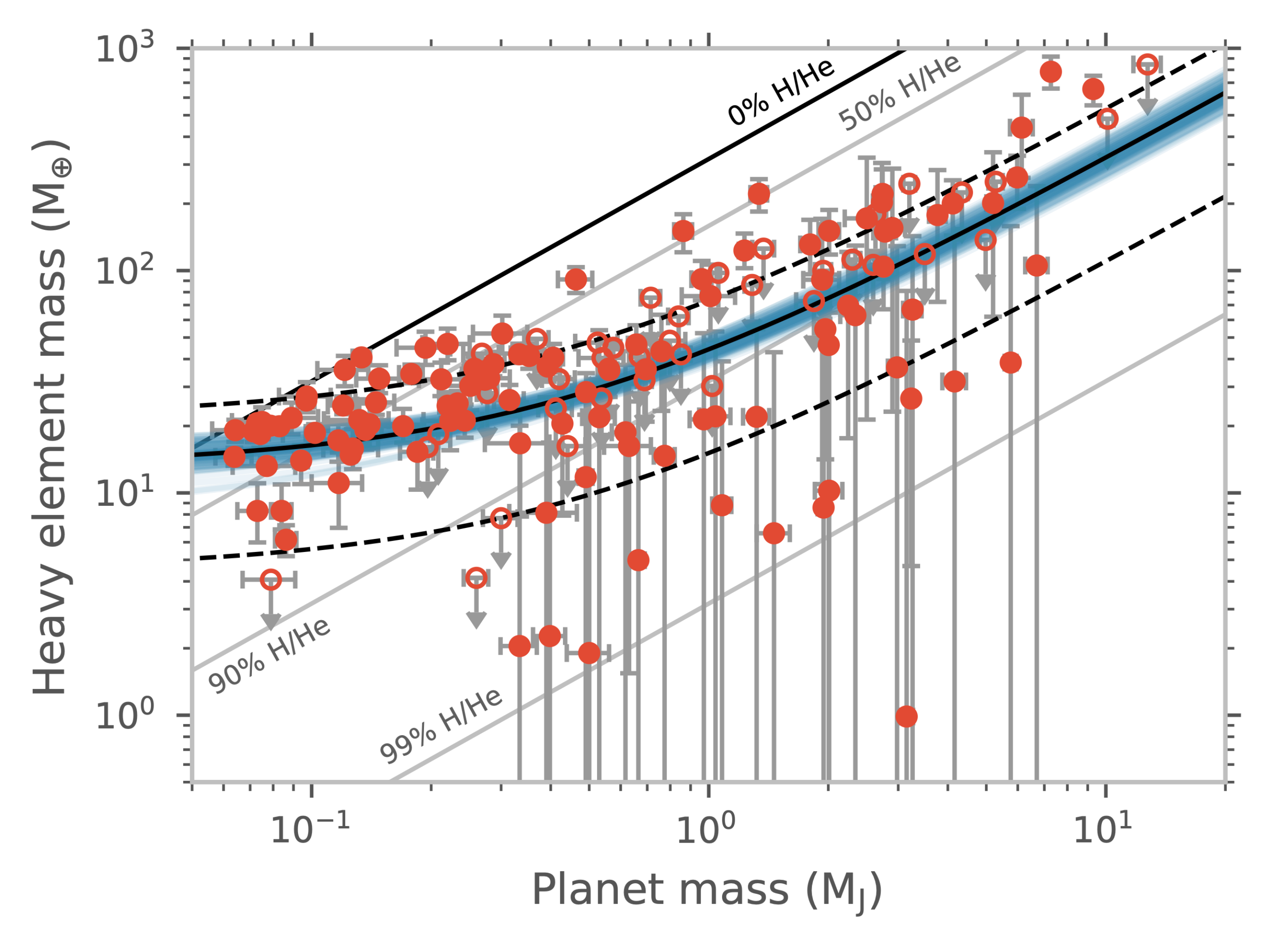}
    \includegraphics[width=0.44\linewidth]{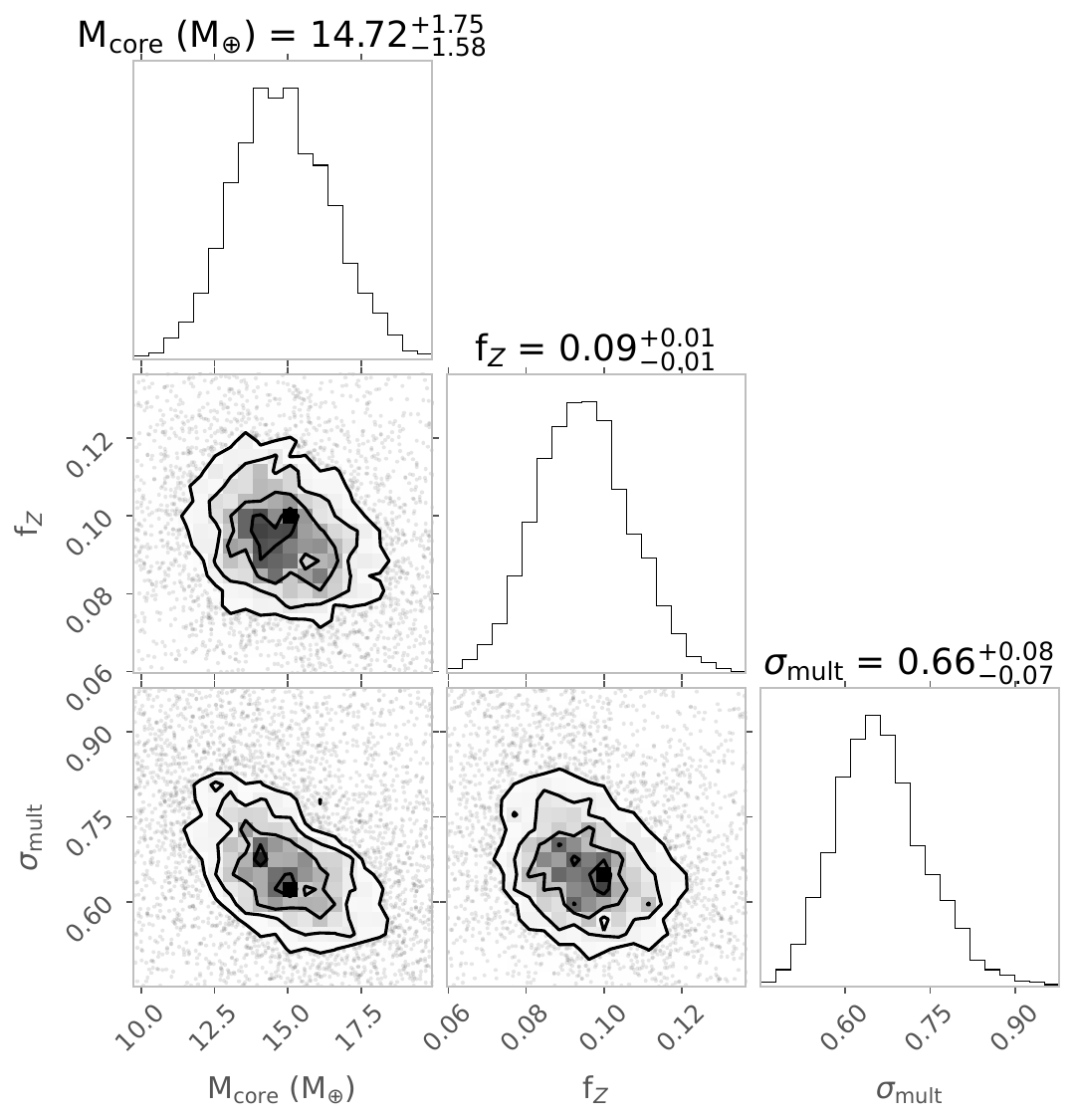}
    \caption{The left panel shows our metal mass measurements against planet mass, a sample of models from our fiducial model in blue, mean of the models in solid black and the range of astrophysical scatter in dashed black lines. Solid straight lines show metal mass as a function of planet mass for a range of metallicities. The right panel shows the constraints obtained on the fiducial model parameters: core mass $M_{\rm core}$, fraction of accreted material in metals $f_Z$, and multiplicative term for the astrophysical scatter $\sigma_{\rm mult}$.}
    \label{fig:fiducial_fit}
\end{figure*}

\textit{Metal poor low mass planets:} Although planets $<0.1$ M$_{\rm Jup}$ in our sample tend to cluster together at $Z_{\rm p} \sim 1$, there are five notable planets that have $Z_{\rm p} < 0.5$ and $M_{\rm p} <0.1$ M$_{\rm Jup}$: Kepler-30 d, TOI-1420 b, TOI-2525 b, TOI-1173 b, and Kepler-9 c. TOI-1420 b is the largest amongst these and we can only place an upper limit on its metal mass. If the flux threshold for inflation depends on planetary mass, it is possible that TOI-1420 b ($T_{\rm eq} = 955$ K) and potentially TOI-1173 b ($T_{\rm eq} = 943$ K) are inflated in size. Given that $Z_{\rm p} \sim 0.5$ is typically seen as the threshold for runaway gas accretion, it is puzzling why these planets ended up with such low bulk metallicities and yet did not accrete more nebular gas to turn into a more massive giant planet. Follow up observations of these `stalled' giants would be helpful in understanding the runaway accretion threshold.

\textit{Planets with discrepant fitted ages:} There are six planets for which the reported ages are essentially unconstrained but fitting the radius pushes their age posteriors to values below 0.5 Gyr. These planets are some of the largest in size in the sample and they straddle the 0.5 Gyr pure H-He mass-radius curve shown in Figure~\ref{fig:planet_sample}. These planets are TOI-3757 b, HATS-47 b, TOI-4914 b, TOI-1130 c, HATS-77 b, TIC 46432937 b. They span a wide mass range from $0.27 - 3.2$ M$_{\rm Jup}$, are cooler than 900 K, and seem to orbit main-sequence stars. Another planet for which  we do not have a reported age in our sample but the fitted age is young ($0.61^{+0.18}_{-0.08}$ Gyr) is WASP-69 b,. We note that WASP-69 b's large size could be a result of radius inflation given its low mass ($0.26$ M$_{\rm Jup}$) and high $T_{\rm eq} = 959$ K. 

\textit{Planets around evolved stars:} 
Three of the planets that lie above the 0.5 Gyr pure H-He mass-radius curve in Figure~\ref{fig:planet_sample} orbit stars evolving off the main-sequence: Kepler-432 b (0.575 M$_{\rm Jup}$, $4.2^{+0.8}_{-1.0}$ Gyr), Kepler-87 b (1.02 M$_{\rm Jup}$, $7.5 \pm 0.5$ Gyr), and TIC 241249530 b (4.98 M$_{\rm Jup}$, $3.2 \pm 0.5$ Gyr). Planets may be re-inflated due to increased incident flux when stars evolve off the main sequence \citep{Lopez2016, Grunblatt2017}. However, these planets have equilibrium temperatures $< 1000$ K, i.e. below the threshold at which radius inflation is known to become significant. Two of the planets have significant eccentricities (Kepler-432 b: $0.478 \pm 0.007$, TIC 241249530 b: $0.941 \pm 0.001$, 
\citealt{Grunblatt2018} note that giant planets around evolved stars are often eccentric) and may be tidally inflated; Kepler-87 b is on a mildly eccentric orbit with $e = 0.036 \pm  0.009$. It is possible that limitations in accurate determination of stellar radii during this evolutionary phase may be responsible for the anomalously large radii of these planets.

\section{The planet mass-metallicity relation}
\label{sec:Mp_Zp_relation}

\begin{figure*}
    \centering
    \includegraphics[width=0.55\linewidth]{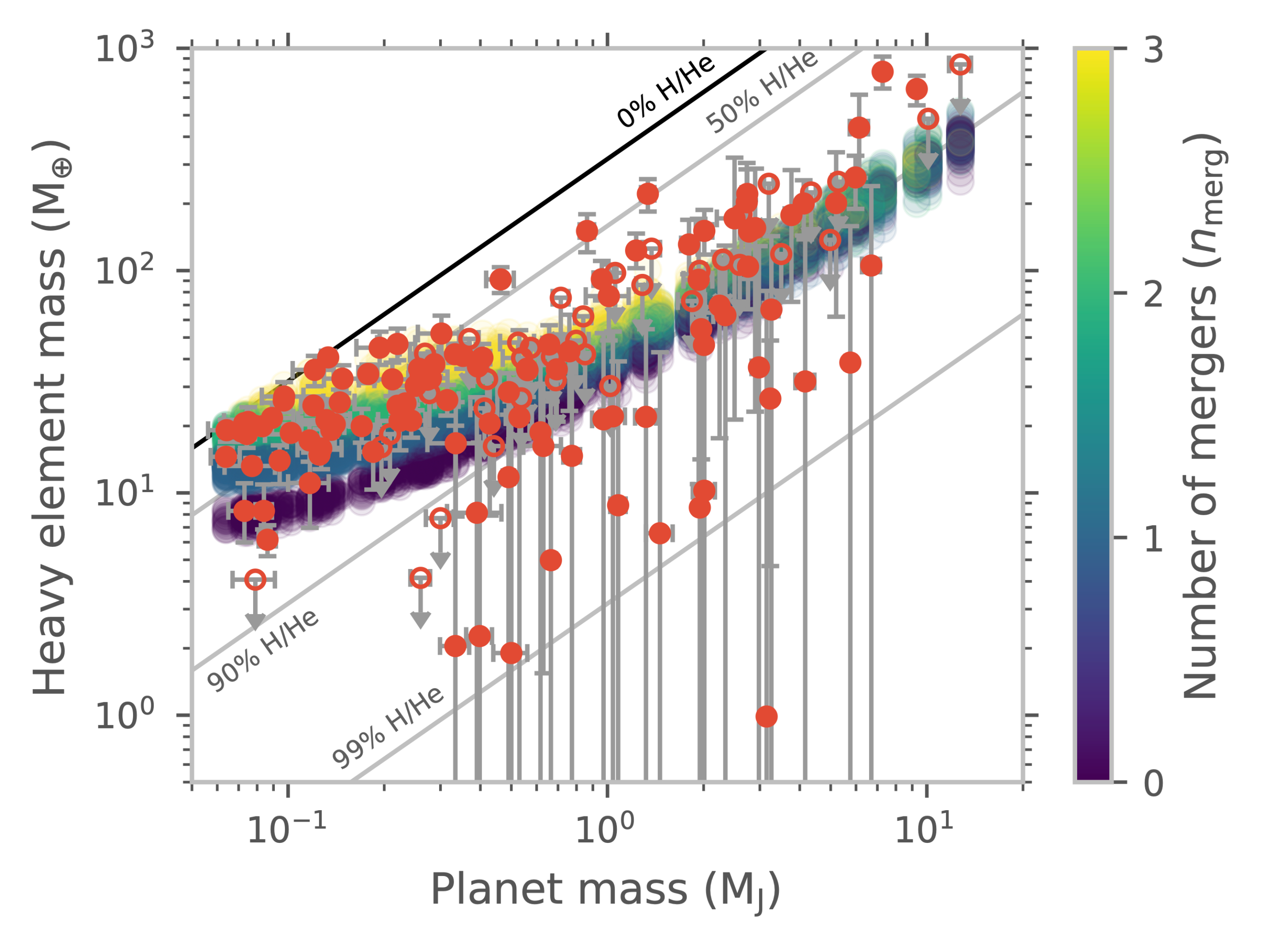}
    \includegraphics[width=0.44\linewidth]{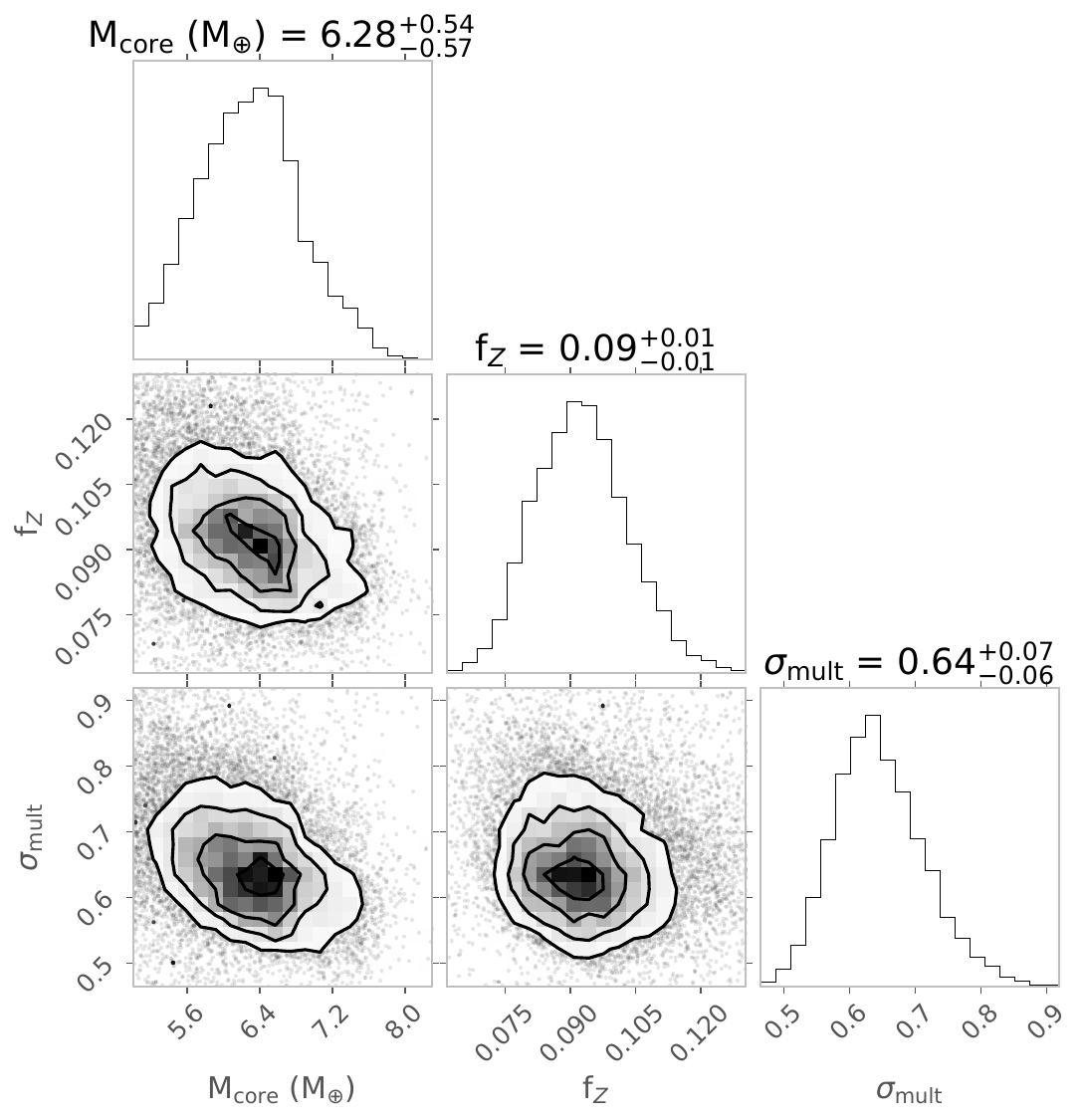}
    \caption{A sample of models and parameter constraints for our mergers model with mean number of mergers $\mu = 1.5$. In the left panel, we color model predictions by the fitted values of $n_{\rm merg}$. Solid straight lines show metal mass as a function of planet mass for a range of metallicities. The fitted core mass $M_{\rm core}$ is roughly half the value from our fiducial model fits. The fraction of accreted material in metals $f_Z$ and astrophysical scatter $\sigma_{\rm mult}$ remain unchanged.}
    \label{fig:mergers_fit}
\end{figure*}

We perform our population-level fits in the metal mass $M_Z$ versus planet mass $M_{\rm p}$ space. For planets with well-constrained metal mass, we use the median and the 1 $\sigma$ uncertainty to calculate the model likelihood in the standard way. For planets with upper limits on their metal masses, we instead use a truncated normal fitted to their $M_Z$ posteriors to calculate the likelihood of the model. The errors on $M_{\rm p}$ are much smaller than the errors on $M_Z$ and also much smaller than the dynamic range in $M_{\rm p}$ so ignoring their errors in our simple fits is acceptable. We verify this by performing the fits with \texttt{linmix} and a custom-built likelihood based on \cite{kellyAspectsMeasurementError2007}, these account for errors on $M_{\rm p}$ in the fit and we find that the results are identical. The simplest model we fit to the population is
\begin{equation}
    \label{eq:fiducial_model}
    M_Z = M_{\rm core} + f_Z (M_{\rm p} - M_{\rm core}).
\end{equation}
Such a model assumes that for $M_{\rm p}$ up to $M_{\rm core}$, the bulk metallicity of the planet $Z_{\rm p} = 1$. The parameter $f_Z$ quantifies the metallicity of the accreted material after $M_{\rm p} > M_{\rm core}$ and can be regarded as the metal mass fraction of the envelope. We find that this model outperforms a simple power law $M_Z = \alpha M_p^{\beta}$ by Bayes factor of 921 ($\Delta$ ln$Z = 6.8$, $Z$ is the Bayesian evidence). This is a significant departure from the mass-metallicity relation found in \cite{Thorngren2016}, where the data did not support a relation more complex than a power law. It is also indicative of how our larger and more precise sample better captures the metallicity distribution of planets at both the low- and high-mass end. Our results are also in qualitative agreement with the inferences of \cite{Muller2020}, who suggested that the bulk metallicity of giant planets is independent of planet mass if one removes $\sim 20$ M$_\oplus$ of metals from each planet in the population. We choose to represent additional scatter as a fraction $\sigma_{\rm mult}$ of the predicted model value $M_Z$ to capture the scatter for a large range of $M_Z$ values. Figure~\ref{fig:fiducial_fit} shows a sample of fitted models and the constraints we obtain on $M_{\rm core}$, $f_Z$, and $\sigma_{\rm mult}$. The core mass $M_{\rm core} = 14.7^{+1.8}_{-1.6}  \, {\rm M}_{\oplus}$ is primarily determined by lower mass planets where $M_Z \sim M_{\rm core}$. The fraction $f_Z$ of incoming mass beyond $M_{\rm core}$ that is in heavy elements is constrained by planets more massive than Jupiter. The contributions to the total metal mass from $M_{\rm core}$ and $f_Z \times M_p$ becomes comparable at $\sim 0.5~M_{\rm Jup}$. Remarkably, the constrained value of $f_Z = 0.09 \pm 0.01$ is well above the solar heavy element mass fraction value of 0.014. We also fit other physically motivated models (\S~\ref{sec:implications}, Appendix \ref{sec:additional_models}) to the data but find that none of them are significantly better at capturing the population-level behavior. We test if the inclusion of low-mass low-metallicity planets that might be potentially inflated (see \S~\ref{sec:bulk_metallicity_estimate}) affects our population level inferences. Refitting the fiducial model to the population while excluding 12 planets with $\leq 0.3$ M$_{\rm J}$ and $T_{\rm eq} > 900$ K, we find that $M_{\rm core} = 15.9 \pm 1.8\,{\rm M}_{\oplus}$ and $f_Z = 0.09 \pm 0.01$ are entirely consistent with our fits to the whole population.

Given the substantial population of $20-40 \, M_\oplus$ planets with $Z_p \sim 1$ and a large constrained value of $M_{\rm core} \sim 15 M_\oplus$, it is surprising that these planets did not accrete more gas. In addition, there is a large variance (factor of $\sim 3$) in the metal mass in this planet mass range, which is notable especially given the small uncertainties in the metal mass. To explore the possibility that such planets form by mergers of lower mass planets \citep{Frelikh2019, Ginzburg2020}, we also fit an additional model to the planet population:
\begin{equation}
    \label{eq:collision_model}
    M_Z = (n_{\rm merg} + 1) M_{\rm core} + f_Z [M_p - (n_{\rm merg} + 1) M_{\rm core}],
\end{equation}
where $n_{\rm merg}$ is the number of mergers a planet undergoes. Such a formalism keeps the planet bulk metallicity constant if it grows by mergers. Since $n_{\rm merg}$ needs to be different for each planet, the number of free parameters that need to be fitted to the planet population proliferate. To make progress, a strong prior needs to be placed on $n_{\rm merg}$. We choose to place a Poisson prior on this parameter with a mean number of mergers $\mu_{\rm merg}$ fixed to values in the set $\{0.5, 1, 1.5, 2\}$. Such a prior makes low values of $n_{\rm merg}$ in the range $0-2$ more likely, with higher values of $n_{\rm merg}$ increasingly less likely. This is what we expect from a formation perspective as mergers increase the spacing between planets and the timescale for mergers is extremely sensitive to this spacing \citep{Pu2015}. Figure~\ref{fig:mergers_fit} shows our fitted models and the constraints on the main parameters of the fit. Even with the addition of a $n_{\rm merg}$ parameter for each planet, we obtain a Bayes factor of $\sim 1$ compared to the fiducial model for $\mu = 1.5$ and $2$. Lower values of $\mu$ are weakly disfavored relative to the fiducial model by the Bayes factor. The fitted value of $M_{\rm core} \sim 6.3 ~M_\oplus$ is now roughly half the value in the fiducial model but $f_Z$ and $\sigma_{\rm mult}$ are unchanged. Given the formulation of our model, variations in $n_{\rm merg}$ mostly affect the predicted $M_Z$ at lower planet masses. More massive ($\gtrsim 1~M_{\rm Jup}$) planets with high $Z_p$ and $M_Z$ continue to not be well fitted by this model and would require very high number of mergers ($\sim 6-15$) of $M_{\rm core}$ mass objects. 

We also test whether the metal mass distribution can be fitted by a $M_{\rm core}$ that grows with $M_{\rm p}$ while $f_Z$ is fixed to the solar value of 0.014. Such a fit yields $M_{\rm core} \propto M_{\rm p}^{0.5}$ and a Bayesian evidence comparable to the one obtained by the power law fit and therefore strongly disfavored relative to our fiducial model. Fixing the power law dependence to $M_{\rm core} \propto M_{\rm p}^{1/5}$ (as suggested for mergers by \citealt{Ginzburg2020}) while letting $f_Z$ vary, we recover the fiducial model's significantly super-stellar $f_Z = 0.08 \pm 0.01$ with Bayesian evidence comparable to our fiducial model ($\Delta$ ln$Z$ $= -0.97$).

\section{Trends with other properties}

We use our large planet sample to search for the relation between metal mass and properties other than planet mass. In particular, we test whether the metal mass of a planet depends on its host star's mass, as reported by \cite{Muller2025}. We find no evidence of such dependence when we fit for an additional term dependent on stellar mass $M_\star$ in our fiducial model (Equation~\ref{eq:fiducial_model}), indicating that any dependence, if present, is weak and of low significance (see Appendix~\ref{sec:residual_correlation} for residuals in $M_Z$ plotted against $M_\star$). We note that even in \cite{Muller2025}, the difference in metallicity of planets around M stars and FGK stars is comparable to the astrophysical scatter in planet metallicity around FGK stars. It is possible that their adopted method for accounting for astrophysical scatter underestimates it. We perform power law fits that do not account for astrophysical scatter and do find this $M_\star$ dependence.
In addition, the fits in this work are performed with the custom method based on \cite{kellyAspectsMeasurementError2007} since the uncertainties on $M_\star$ are large enough compared to the dynamic range of $M_\star$ for their effect on the model fitting to matter. We do not separate the population by stellar mass to do independent fits on each population.
We also find that fitting $M_Z$ with power laws in $M_{\rm p}$ and $M_\star$ yields a weak $M_\star$ dependence (at $<1~\sigma$ level) but this is due to the fact that a power law in $M_{\rm p}$ does not capture the $M_Z$ dependence on $M_{\rm p}$ (see \S~\ref{sec:Mp_Zp_relation}) and that $M_\star$ and $M_{\rm p}$ are weakly correlated\footnote{We use Gaussian deconvolution (as recommended by \citealt{steinhardtNonparametricMethodsAstronomy2018}) to determine the correlation between log($M_{\rm p}$) and log($M_\star$) while accounting for the uncertainties in these quantities following the methods outlined in \cite{Bovy2011}. Fitting a single 2D Gaussian to the distribution, we find a weak correlation coefficient of $0.19 \pm 0.08$.}. 

\begin{figure}
    \centering
    \includegraphics[width=\linewidth]{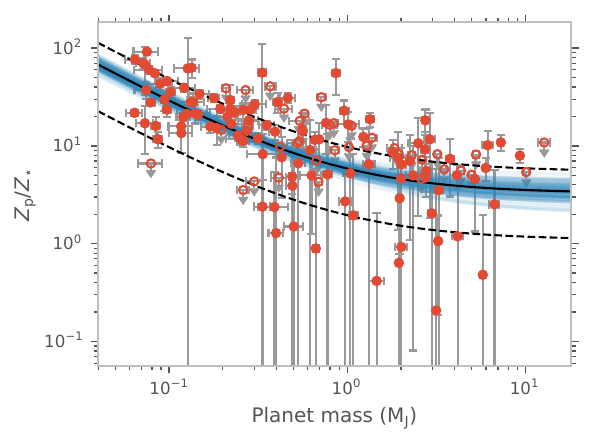}
    \caption{Bulk metallicity normalized by the stellar metallicity plotted against planet mass. Normalizing $Z_{\rm p}$ by $Z_\star$ does not reduce the astrophysical scatter in our sample. Blue lines show fitted samples from our model of the form $Z_{\rm p}/Z_\star = a / M_{\rm p} + b$, where $a = 2.6 \pm 0.3$ (for $M_{\rm p}$ in Jupiter masses) and $b = 3.3 \pm 0.5$. The dashed black lines mark the range of the astrophysical scatter. A noteworthy difference between $Z_{\rm p}/Z_\star$ and $Z_{\rm p}$ in Figure~\ref{fig:planet_metallicity} is that outliers in $Z_{\rm p}$ are less salient in $Z_{\rm p}/Z_\star$ space.}
    \label{fig:ZpZstar_vs_Mp}
\end{figure}

We find a similar lack of dependence of $Z_{\rm p}$ and $M_Z$ on $Z_\star$ (Appendix~\ref{sec:residual_correlation}), in agreement with \cite{teskeMetalrichStarsMake2019}. We also assess the utility of normalizing the bulk metallicities by stellar metallicities as \cite{Thorngren2016} found that the astrophysical scatter in $Z_{\rm p}/Z_\star$ was lower than the one in $Z_{\rm p}$. In Figure~\ref{fig:ZpZstar_vs_Mp}, we show bulk metallicity normalized by stellar metallicity. We assign solar metallicity and the standard deviation of the stellar metallicity in our sample to stars that do not have a metallicity value and metallicity uncertainty, respectively. When we fit a simple model\footnote{We find that this model is strongly preferred to a power law model, in agreement with our conclusions for both $M_Z$ and $Z_{\rm p}$.} of the form in Equation~\ref{eq:fiducial_model}, $Z_{\rm p}/Z_\star = a / M_{\rm p} + b$, we find that $a = 2.6 \pm 0.3$ (for $M_{\rm p}$ in $M_{\rm Jup}$) and $b = 3.3 \pm 0.5$. Taking median values of $a$ and $b$, we would expect 0.1, 1, and 10 M$_{\rm Jup}$ mass objects to be enriched by 
a factor of $\sim 29.3$ , 5.9, and 3.6 relative to their host stars. We find that normalization by $Z_\star$ does not reduce the astrophysical scatter ($\sigma_{\rm mult} = 0.67^{+0.08}_{-0.07}$) in our much larger sample. We tested this on a smaller sample of planet hosts with stellar metallicities measured by \emph{Gaia} and came to the same conclusion. However, normalizing $Z_{\rm p}$ by $Z_\star$ does make the high metallicity outlier planets less salient. We note that due to a weak correlation in the sample between log($M_\star$) and log($Z_\star$) ($-0.18 \pm 0.08$, quantified using Gaussian deconvolution, \citealt{Bovy2011}), normalizing $Z_{\rm p}$ by $Z_\star$ can introduce spurious $M_\star$ dependence in $Z_{\rm p}$. We also test if the metal mass depends on the orbital period, orbital eccentricity, and equilibrium temperature of the planets but find no evidence in support of such hypotheses (Appendix~\ref{sec:residual_correlation}).

\section{Implications and conclusions}
\label{sec:implications}

Our measurements of a large sample of giant planets across a wide mass range reveal key insights regarding the formation of giant planets. Firstly, they suggest that $M_{\rm core}$ is in the range of $6 - 15~M_\oplus$, depending on whether mergers played an important role in the formation of hot super-Neptunes. This range of $M_{\rm core}$ is consistent with the most massive volatile-poor super-Earths that we find \citep[e.g.,][]{Dai2021}. The upper limit is also consistent with the estimated metal mass of Neptune and Uranus in our solar system \citep{Nettelmann2013, Helled2020}. We note that the astrophysical scatter in $M_Z$ corresponds to scatter in $M_{\rm core}$ for this planet mass range so core masses in the fiducial model could range from $5 - 23 ~M_\oplus$. A majority of the planets $< 30 \, M_\oplus$ in our sample only accreted small amounts of nebular gas and have $Z_p \gtrsim 0.7$. 

\begin{figure}
    \centering
    \includegraphics[width=\linewidth]{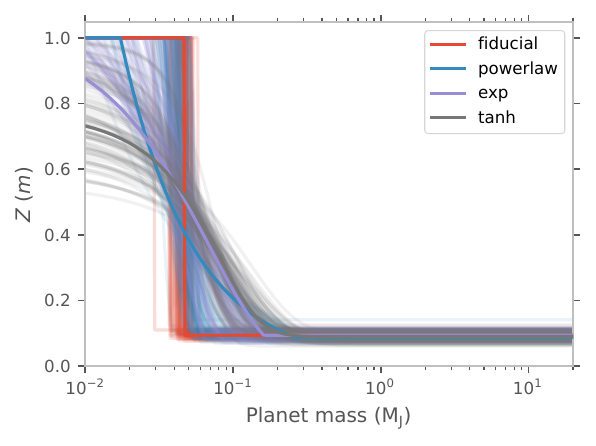}
    \caption{Fitted profiles for the metallicity of the incoming material $Z(m)$ as a function of planet mass $m$.}
    \label{fig:Zm_profiles}
\end{figure}

Secondly, the planets in our sample reach $Z_{\rm p} = 0.5$ when they are $\sim 30 - 60~M_\oplus$. The classical paradigm, which predicts $Z_{\rm p} \sim 1$ at $10~M_\oplus$ and $Z_{\rm p} \sim 0.5$ at $20~M_\oplus$ \citep[e.g.,][]{Mizuno1980, Pollack1996}, is inconsistent with the population. Even in the $60 - 90~M_\oplus$ range, 13 out of 18 planets have $Z_{\rm p}> 0.25$, which is comparable to the bulk metallicity ($Z_{\rm p} = 0.2$) of Saturn. A handful of planets are exceptions to this population-level behavior and exhibit lower $Z_{\rm p} \sim 0.1 - 0.3$ even when they are in the $20-30~M_\oplus$ mass range. Figure~\ref{fig:Zm_profiles} shows the metallicity of the incoming material $Z(m)$ as a planet grows in mass for our simple formation models (fiducial as well as additional models in Appendix~\ref{sec:additional_models}). For the fiducial model, the cumulative $Z_{\rm p} = \int Z(m) dm / M_{\rm p} = 0.5$ that is considered the threshold for runaway accretion is reached when planet mass is $32 \pm 3 \, M_\oplus$. Similar constraints are obtained for the other models, which probe if there is any observational evidence for a gradual decline in $Z(m)$. Such models provide comparable fits to the population as the fiducial model so a detection of a gradient in $Z(m)$ remains elusive. However, the fitted $Z(m)$ profiles decline to $Z_{\rm min}$ pretty rapidly, typically in the mass range of $30-40 \, M_\oplus$ (see Appendix~\ref{sec:additional_models}). Although more protracted than the classical accretion scenario, the fitted models likely still fall short of the extended accretion needed to create the dilute cores of the solar system giants \citep{Helled2024}.

\begin{figure}
    \centering
    \includegraphics[width=\linewidth]{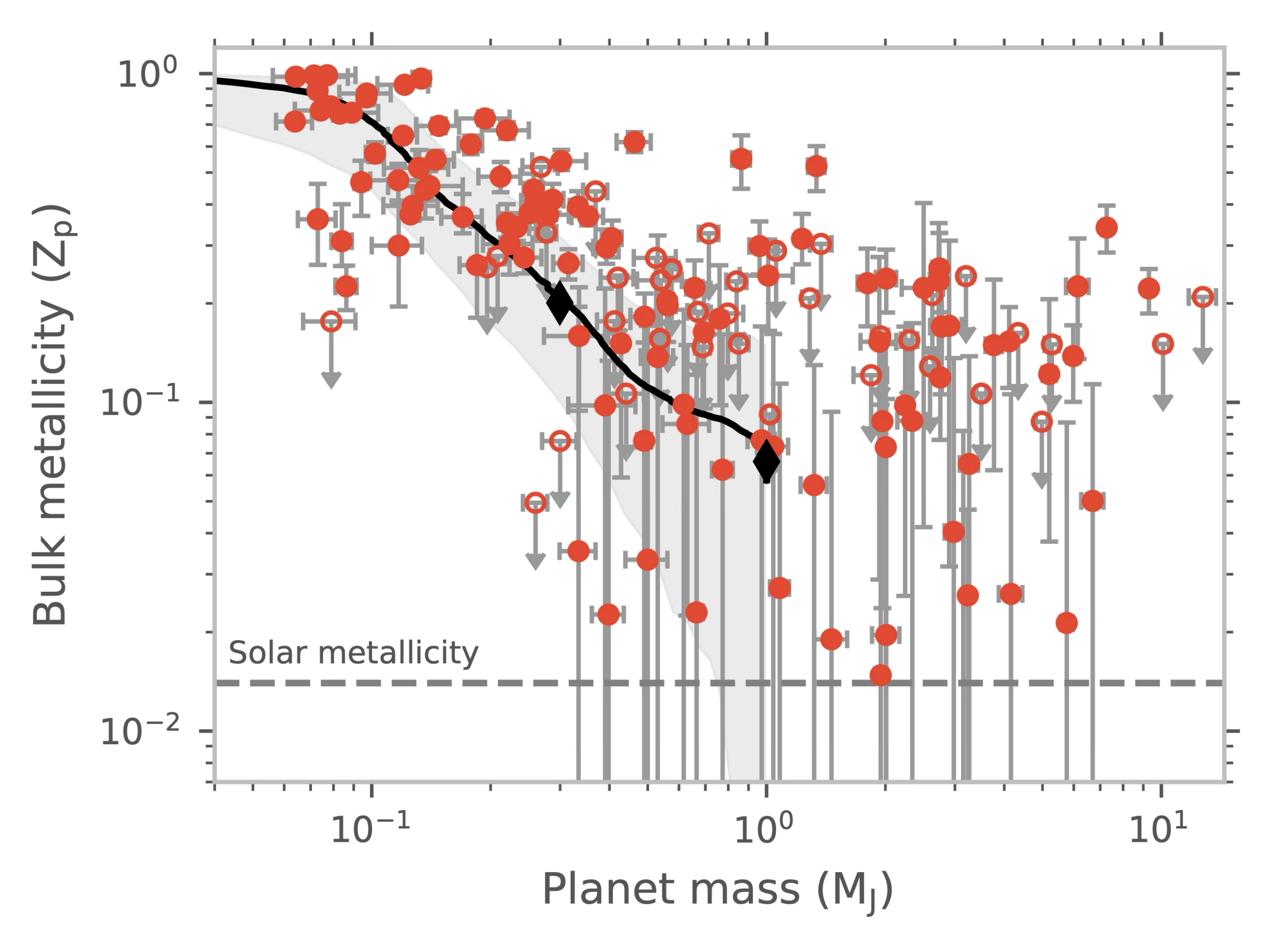}
    \caption{Our bulk metallicity measurements with the formation sequence from \cite{Helled2023} (only available upto $1~M_{\rm Jup}$.) overlaid on the population. The black line shows the median outcome and the gray region encompasses the range of formation tracks. The dashed horizontal line indicates solar metallicity $Z_\odot = 0.014$. The black diamonds show the bulk metallicity estimates for Saturn \citep[$M_Z = 19.1 \pm 1.0 \, M_\oplus$;][]{Mankovich2021} and Jupiter \citep[$M_Z \sim 21 \pm 3 \, M_\oplus$ for unmodified EOS;][]{Howard2023a}.}
    \label{fig:Helled-track}
\end{figure}

The large inferred $M_{\rm core}$, high metallicities of $< 30$ M$_\oplus$ objects, and $Z_{\rm p}$ reaching 0.5 at $\sim 30 - 60~M_\oplus$ suggests that runaway accretion is delayed relative to classical expectations. This could be a result of gas accretion slowing down due to protracted solid accretion \citep[e.g.,][]{Alibert2018, Helled2023}, the recycling of high entropy disk gas through a growing planet's Hill sphere \citep{Ormel2015, Cimerman2017, Ali-Dib2020}, or elevated opacities from the dust grains that radially drift to the inner disk \citep{Chachan2021}. 
In Figure~\ref{fig:Helled-track}, we show formation tracks from \cite{Helled2023}, who argue that Saturn may be a `failed' giant and that the runaway stage is delayed. The agreement between the models and the planet population is striking. Alternatively, mergers of lower mass planets with high $Z_{\rm p}$ during or after the dissipation of the gas disk could also preserve or even increase (if envelope loss ensues) the bulk metallicity of a planet while increasing its mass \citep{Ikoma2006, Ginzburg2020, Chachan2025}. However, we do not see a significant correlation between planet eccentricity and metallicity so either mergers are not important or the post-merger eccentricity is damped out by dynamical friction from local solid or gaseous material.

Thirdly, as the planet mass increases, the population's bulk metallicity plateaus at significantly super-solar values ($\sim 7 \times$ solar) instead of continuing to decline to lower and lower values. This is a remarkable departure from expectations from a power-law like relationship between $M_{\rm p}$ and $Z_{\rm p}$. Even after taking the astrophysical scatter in account, we expect $Z_{\rm p}$ for most planets to be in the range of $[0.33, 1.66] \times 0.09 \sim 0.03 - 0.15$, which is roughly equivalent to $2 - 12 \times$ solar. Jupiter's atmospheric metallicity of $\sim 3 \times$ solar is in agreement with this population-wide trend. These elevated bulk metallicities reveal giant planets continue to preferentially accrete metals as they grow in mass. This metallicity enhancement could be due to accretion during later stages of disk evolution when preferential loss of nebular hydrogen and helium enriches the gas left behind \citep{guillotCompositionJupiterSign2006, deschJupitersNobleGas2014}. Alternatively, large scale pebble drift and evaporation may enrich the local gas in volatiles, which is then accreted by these planets \citep{Schneider2021, Ohno2025}. In contrast to these scenarios, giant planets may also be enriched during the runaway phase via solid accretion. This could be in the form of planetesimals ($\gtrsim 0.1 - 1$ km) that are dynamically decoupled from the gas. These planetesimals interact gravitationally with the growing planet and can be accreted efficiently, especially if the escape velocity from the planet ($\sim$ size of the dynamical kick) is smaller than the local orbital velocity and if their dynamical excitation can be damped by ambient disk gas \citep{Guillot2000, Goldreich2004a}. These conditions are met more easily in the inner disk in the regions where the transiting planets in our planet sample are found. In the outer disk, planetesimal accretion may be inefficient due to increased likelihood of ejections.  However, the extent of accretion depends sensitively on the assumed rate of damping of this excitation by ambient gas as well as the accretion cross-section of the planet \citep{Zhou2007, Shiraishi2008, Shibata2019, Eriksson2022}. A large amount of planetesimals could also potentially be accreted if giant planets migrate large distances through their natal disks to reach their current locations \citep{Shibata2020}. Planets could alternatively accrete smaller solids that are better coupled to the gas \citep{Morbidelli2023, Bitsch2023}. Grains that are mm-cm sized with Stokes number $\sim 0.1 - 1$ may get trapped in disk substructures carved by a giant planet (though the simulations in 3D of flow structures through the gap suggest that trapping may be inefficient, \citealt{Huang2025}) but smaller grains pass through these traps \citep{Drazkowska2019, Stammler2023}. Preliminary studies show that the accretion efficiency of these grains is small \citep{VanClepper2025}. A better understanding of their dynamics, especially in the vicinity of the planet, would help evaluate the feasibility of metal enrichment via the accretion of such solids. Whether giant planets are enriched in metals due to accretion  of solids or metal-enriched gas could potentially be determined by quantifying the refractory and volatile composition of these planetary envelopes \citep{Lothringer2021, Turrini2021, Chachan2023}. 

Given the orbital distance dependence of the various metal accretion processes discussed here, our measurements may place additional useful constraints on the migration history of close-in giant planets at the population level. Formation at larger orbital distances, for example in regions where giant planet occurrence rate peaks \citep{Fernandes2019, Fulton2021}, followed by high eccentricity migration versus formation concurrent with substantial disk migration may produce different levels of expected metal enrichment. Alternatively, metal accretion close to the planets' current locations may be setting their bulk metallicities. More detailed formation modeling is necessary to determine if the bulk metallicity distribution can distinguish between these scenarios and we leave this endeavor to future work.

Overall, the transiting giant planet population exhibits multiple properties that support the core accretion paradigm of planet formation. The planets are metal-dominated when $<30$ M$_\oplus$ and the inferred core mass is in the range of $6-15$ M$_\oplus$, depending on whether mergers are important or not. They only reach the formal threshold for runaway accretion ($Z_{\rm p} \sim 0.5$) in the $30-60$ M$_\oplus$ range and they continue to preferentially accrete metals ($f_Z = 0.09 \pm 0.01$) as they grow into Jupiters and super-Jupiters. We do not find evidence for any dependence of bulk metallicity on additional planetary and stellar properties. 

These results also offer a benchmark for determining the provenance of distant giant planets. If distant giants form by gravitational instability, we would not expect them to be as enriched in metals in their bulk as the transiting giant planets (\citealt{Helled2010}, though modest enrichment is possible \citealt{Boley2011}). If planets accrete metals primarily through planetesimals, the distant giants' enrichment may also be lower because planetesimals are more likely to get ejected at large distances (\citealt{Goldreich2004a}, see earlier discussion of other factors that affect planetesimal accretion). Empirical measurements of the bulk metallicity of distant giants and a comparison with transiting planets would therefore be extremely useful for understanding how giant planets accrete heavy elements. Although currently both the masses and radii of distant giant planets are determined indirectly, astrometry and radial velocity measurements will yield dynamical masses for these planets and enable tighter constraints on their bulk metal abundances in the future. In the meantime, atmospheric metallicity can serve as a useful probe of a giant planet's metal content as it likely represents a lower limit on their bulk metallicity (\citealt{Thorngren2019}, see \citealt{Debras2019, Debras2021, Howard2023c} for developments about potential differences in Jupiter's envelope and atmospheric metallicities). Ongoing and future surveys of atmospheric composition of distant giant planets will provide initial insights into giant planet formation across a wide range of orbital distances.

\section*{Acknowledgments}
We thank the referee for asking thought-provoking questions that strengthened the conclusions in this paper. This research has made use of the NASA Exoplanet Archive, which is operated by the California Institute of Technology, under contract with the National Aeronautics and Space Administration under the Exoplanet Exploration Program. Y.C. is thankful to Aida Behmard for guidance in using the \emph{Gaia} metallicities as part of this analysis, to Artem Aguichine for pointing him to the correct entropies for the \cite{Mazevet2019} water EOS, and to Yao Tang for recommending the use of entropy in $k_{\rm B}/$atom. Y. C. is grateful to Sarah Blunt, J. B. Ruffio, Jerry Xuan, and J. J. Zanazzi for thoughtful conversations. We also thank Ravit Helled and Henrik Knierim for helpful exchanges regarding their recent work. J.J.F. and Y.C. acknowledge the support of NASA XRP grant 80NSSC24K0691.

\vspace{5mm}

\software{astropy \citep{2013A&A...558A..33A, 2018AJ....156..123A},  dynesty \citep{Speagle2020}}

\bibliography{Mass-Metallicity}

\appendix
\section{Mixing entropy estimate}
\label{sec:S_mix_estimate}
\begin{figure}
    \centering
    \includegraphics[width=0.5\linewidth]{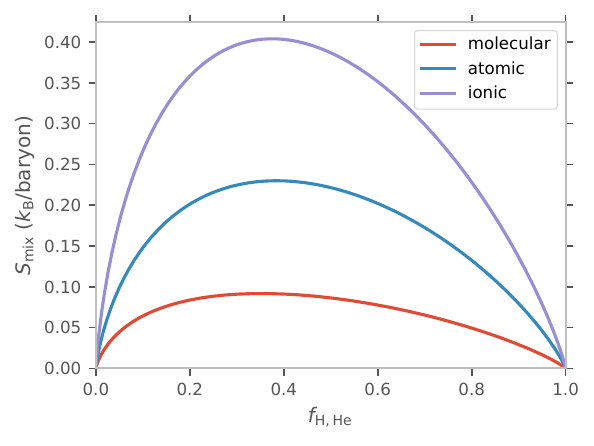}
    \caption{Mixing entropy that arises as a result of mixing water and H-He as a function of the H-He mass fraction ($Y = 0.275$). The different lines correspond to regions in a planet's interior where H-He and water exist in different forms (molecular, atomic, and ionic).}
    \label{fig:Smix_estimate}
\end{figure}

We estimate the entropy arising from mixing of water and H-He in different regimes that are characterized by different degree of dissociation of the planetary constituents. The entropy of mixing is given by
\begin{equation}
    s_{\rm mix} / k_{\rm B} = \sum_{\rm n} - \chi_{\rm n} ~ {\rm ln}(\chi_{\rm n}),
\end{equation}
where $\chi_{\rm n}$ is the number fraction of the n$^{\rm th}$ species. We divide $s_{\rm mix} / k_{\rm B}$ by the mean molecular weight $\mu$ of the mixture to obtain the mixing entropy in $k_{\rm B}$/baryon. The number fraction of each species is given by $ \chi_{\rm n} = (\mu / \mu_{\rm n}) X_{\rm n}$, where $X_{\rm n}$ is the mass fraction and $\mu_{\rm n}$ is the mean molecular weight of the species. We limit ourselves to the fully molecular, fully atomic, and ionic regimes (electron loss only from H) as it is simple to estimate the mean molecular weight and number fraction in these regimes. In the ionic regime, we ignore the effect of indistinguishability of electrons on our estimate of mixing entropy. Figure~\ref{fig:Smix_estimate} shows the mixing entropy of a mixture of H-He and water as a function of the H-He mass fraction ($f_{\rm H, He}$) in these regimes. At a fixed $f_{\rm H, He}$, $s_{\rm mix}$ rises as ones goes from the outer molecular regions to deeper atomic and ionized regions of a planet. This increase would make the adiabats slightly shallower, i.e. the temperature would rise a little less steeply with pressure if $s_{\rm mix}$ is taken into account.  

\section{Additional Model Fits}
\label{sec:additional_models}

Along with our simplest models, we also test if the planet metallicity data are well fitted by other physically motivated models. We assume different functional forms for the metallicity of the accreted material $Z (m)$ when $M_{\rm p} = m$ and calculate $M_Z = \int_0^{M_{\rm p}} Z(m) {\rm d}m$ as a function of $M_{\rm p}$. In our fiducial model, $Z (m) = 1$ for $m \leq M_{\rm core}$ and $Z (m) = f_Z$ for $m > M_{\rm core}$. The models explored below provide different ways of parameterizing how $Z(m)$ changes with $M_{\rm p}$ and with the addition a single free parameter, enable us to probe hints of a gradient in $Z(m)$. We consider a model in which $Z (m)$ has a power law dependence on $m$:
\begin{align}
&Z(m)= 
&\begin{cases} 
    1 & m \leq M_{\rm core}\\
    (m/M_{\rm core})^{-\alpha} & M_{\rm core} 
    < m \leq M_{Z_{\rm min}} \\
    Z_{\rm min} & m > M_{Z_{\rm min}},
\end{cases}
\label{eq:Zm_powerlaw}
\end{align} 
where $M_{Z_{\rm min}} = Z_{\rm min}^{-1/\alpha} M_{\rm core}$. We also consider a model in which $Z(m)$ in the intermediate mass regime falls off exponentially with a characteristic width $\sigma_M$ instead:
\begin{equation}
    Z(m)= {\rm exp}[-(m - M_{\rm core}) / \sigma_M]  \hspace{0.5cm} M_{\rm core} 
    < m \leq M_{Z_{\rm min}} 
    \label{eq:Zm_exp}
\end{equation}
and $M_{Z_{\rm min}} = -{\rm ln}(Z_{\rm min}) \sigma_M + M_{\rm core}$. The third model that we employ to fit the data utilizes a hyperbolic tangent function to model $Z(m)$:
\begin{equation}
    Z(m)= a \; {\rm tanh}[b (m - c)] + e,
    \label{eq:Zm_tanh}
\end{equation}
where $b$ quantifies how rapidly $Z(m)$ changes, $c$ quantifies the planet mass at which $Z(m)$ changes most rapidly (and $Z(m) = e$ for $m = c$), $a = (Z_{\rm min} - 1) / 2$ and $e = (Z_{\rm min} + 1) / 2$. 

We find that the fitted parameters for all of these models yield $M_Z - M_{\rm p}$ relations that closely match the one obtained from our simplest fiducial model. Importantly, the retrieved $Z_{\rm min}$ is identical to the value we report. In general, the fitted $Z(m)$ profiles rapidly decline to the $Z_{\rm min}$ value: $M_{Z_{\rm min}} = 21^{+16}_{-4}~M_\oplus$ for the power law model, $M_{Z_{\rm min}} = 38^{+12}_{-14}~M_\oplus$ for the exponential model, and $b~(m-c) = 1$ for $m = 35^{+10}_{-9}~M_\oplus$ for the hyperbolic tangent model. The Bayes factor obtained from these fits is $\sim 1$ ($\Delta {\rm ln}(Z) = 0.69, -0.19, -0.39$ for the power law, exponential, and hyperbolic tangent models relative to the fiducial model, respectively) so none of these models are significantly better at explaining the data and evidence for a gradient in $Z(m)$ remains elusive.

\section{Residuals and Correlation Plots}
\label{sec:residual_correlation}

\begin{figure*}
    \centering
    \includegraphics[width=0.49\linewidth]{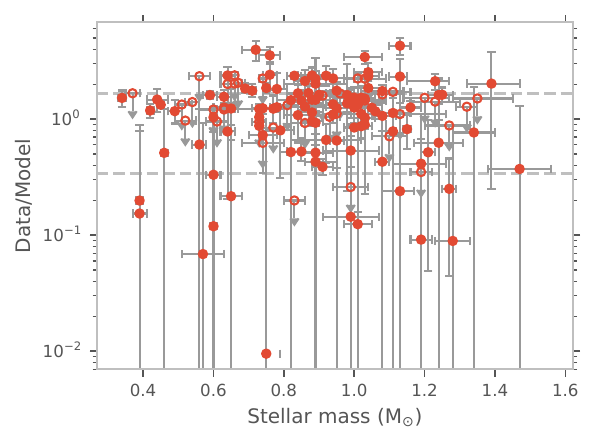}
    \includegraphics[width=0.49\linewidth]{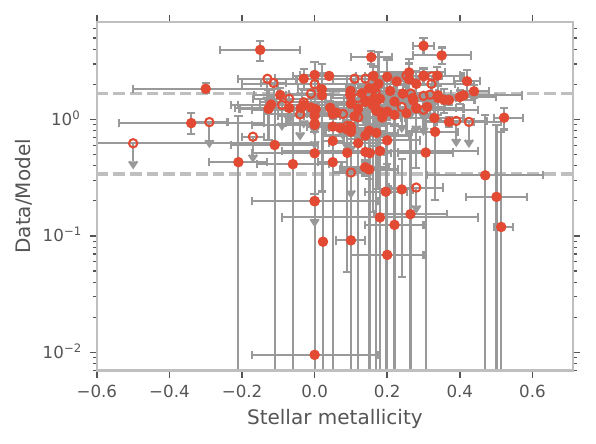}
    \includegraphics[width=0.49\linewidth]{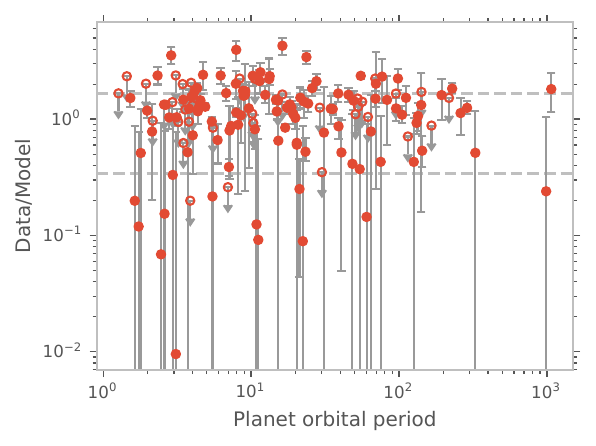}
    \includegraphics[width=0.49\linewidth]{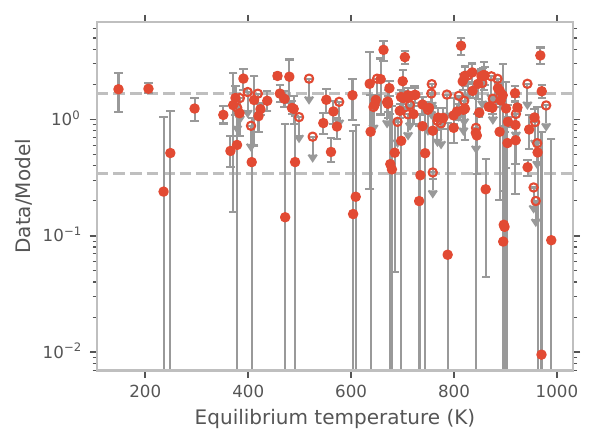}
    \includegraphics[width=0.49\linewidth]{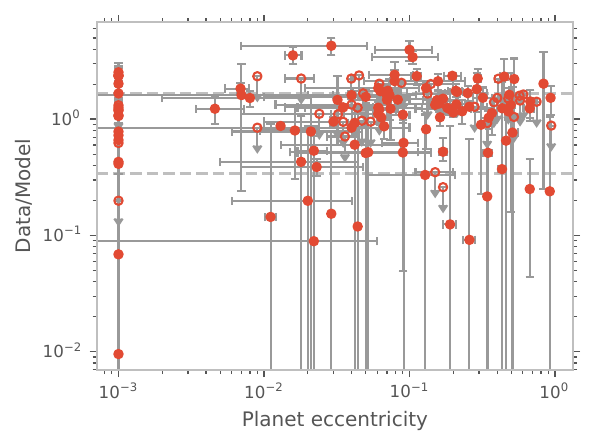}
    
    \caption{Metal mass $M_Z$ divided by the median fiducial model with $M_{\rm core} = 14.72$ M$_\oplus$ and $f_Z = 0.09$ plotted against various stellar and planetary properties. Additional uncertainty from variation in modeled values is not included. The horizontal dashed lines show the range of astrophysical scatter ($1 \pm \sigma_{\rm mult}$ with $\sigma_{\rm mult} = 0.66$). Planets with eccentricity values that are missing or equal to zero are assigned an eccentricity of $10^{-3}$ to display them in log-scale. No correlations between these additional properties and the metal mass residuals are apparent.}
    \label{fig:residual_correlation}
\end{figure*}

Figure~\ref{fig:residual_correlation} shows the residuals (data/model) for metal mass $M_Z$ against various stellar and planetary properties.

\end{document}